\documentclass[12pt]{iopart}
\usepackage[utf8]{inputenc}
\usepackage{amsfonts} 
\usepackage{url} 
\usepackage{cite}
\usepackage{graphicx}

\expandafter\let\csname equation*\endcsname\relax
\expandafter\let\csname endequation*\endcsname\relax

\usepackage{amsmath}
\usepackage{amsthm}
\usepackage[english]{babel}
\usepackage{caption}
\usepackage{subcaption}
\usepackage[table]{xcolor}
\usepackage[ruled,vlined]{algorithm2e}
\usepackage{tablefootnote}
\usepackage{amssymb}
\usepackage{bbm}
\usepackage{bm}

\usepackage{tikz} 
\usetikzlibrary {arrows.meta} 
\usetikzlibrary{shapes.arrows}
\usetikzlibrary{decorations.markings} 

\DeclareMathOperator{\R}{\mathbb{R}}
\DeclareMathOperator{\M}{\mathbb{M}}

\DeclareUnicodeCharacter{2212}{-}

\newcommand{\bnu}{\mbox{\boldmath$\nu$}}

\newcommand{\iu}{\boldsymbol{\mathrm{i}}}

\newcommand{\hA}{\mbox{$\mathcal{A}$}}

\newcommand{\hS}{\mbox{$\mathcal{S}$}}

\newcommand{\hL}{\mbox{$\mathcal{L}$}}

\newtheorem{definition}{Definition}[section]
\newtheorem{theorem}{Theorem}[section]
\newtheorem{lemma}{Lemma}[section]
\newtheorem{proposition}{\bf Proposition}[section]

\newtheorem{corollary}{\bf Corollary}[section]

\newtheorem{example}{Example}[section]

\definecolor{brilliantrose}{rgb}{1.0, 0.33, 0.64}
\definecolor{myviolet}{rgb}{0.21, 0.0, 0.85}
\definecolor{amethyst}{rgb}{0.6, 0.4, 0.8}
\definecolor{carrotorange}{rgb}{0.93, 0.57, 0.13}

\def\bc{\begin{center}}
\def\ec{\end{center}}
\def\bea{\begin{eqnarray}}
\def\eea{\end{eqnarray}}

\date{}

\begin{document}
\title[Higher-order Connection Laplacians for Directed Simplicial Complexes]{Higher-order Connection Laplacians for Directed Simplicial Complexes}
\author{Xue Gong$^{1,2,3,4}$, Desmond J. Higham$^{1,2}$,
Konstantinos Zygalakis$^{1,2}$, and Ginestra Bianconi$^{3,5}$}
\address{$^1$ School of Mathematics, University of Edinburgh, Edinburgh, EH9 3FD, UK}
\address{$^2$ The Maxwell Institute for Mathematical Sciences, Edinburgh, EH8 9BT, UK}
\address{$^3$ The Alan Turing Institute, 96 Euston Rd, London, NW1 2DB, UK}
\address{$^4$ Division of Mathematics, Nanyang Technological University, 50 Nanyang Avenue, Singapore 639798}
\address{$^5$ School of Mathematical Sciences, Queen Mary University of London, London, E1 4NS, UK}
\ead{g.bianconi@qmul.ac.uk}

\begin{abstract}
Higher-order networks encode the many-body interactions existing in complex systems, such as the brain, protein complexes, and social interactions.  Simplicial complexes are higher-order networks that allow a comprehensive investigation of the interplay between topology and dynamics. However, simplicial complexes have the limitation that they only capture undirected higher-order interactions while in real-world scenarios,   often there is a need to introduce the direction of simplices, extending the popular notion of direction of edges. On graphs and networks the Magnetic Laplacian, a special case of Connection Laplacian, is becoming a popular operator to treat edge directionality. Here we tackle the challenge of treating directional simplicial complexes by formulating Higher-order Connection Laplacians taking into account the configurations induced by the simplices' directions. Specifically, we define all the Connection Laplacians of directed simplicial complexes of dimension two and we discuss the induced higher-order diffusion dynamics by considering instructive synthetic examples of simplicial complexes. The proposed higher-order diffusion processes can be adopted in real scenarios when we want to consider higher-order diffusion displaying non-trivial frustration effects due to conflicting directionalities of the incident simplices.
\end{abstract}

\vspace{2pc}
\noindent{\it Keywords}: Connection Laplacian, Simplicial Complex, Higher-order diffusion, Hodge Laplacian, Magnetic Laplacian

\section{Introduction}
Higher-order networks
\cite{battiston2020networks,bick2021higher,torres2021and,salnikov2018simplicial} are attracting increasing attention as they have the ability to encode for the interactions \cite{battiston2021physics} among two or more nodes of complex systems and to display a rich interplay between topology and dynamics \cite{bianconi2021higher,majhi2022dynamics}.  Higher-order networks include hypergraphs \cite{tudisco2021node,gong2023generative,sun2021higher,jost2019hypergraph,mulas2020coupled} as well as simplicial complexes. Simplicial complexes are higher-order networks that are amenable to a comprehensive higher-order topological treatment revealing the topology of data \cite{carlsson2009topology,vaccarino2022persistent,otter2017roadmap,ghrist2008barcodes,meng2020weighted,chazal2021introduction} leading to applications in neuroscience \cite{giusti2016two,petri2014homological,reimann2017cliques}, biology \cite{cang2015topological, chan2013topology, nanda2013simplicial, xia2014persistent}, sensor networks \cite{de2007coverage, ghrist2005coverage}, and computer graphics \cite{Boksebeld2022highorder, Zhao2019Hodge}. Moreover, the algebraic topology of simplicial complexes is drastically changing our understanding of the dynamical state of simple and higher-order networks. Until recently the dynamical description of the network has taken almost exclusively a node (vertex)-centric approach where only the nodes are associated with dynamical variables. The investigation of the dynamical state of simplicial complexes has instead revealed that this is only a special case and that in general each simplex (higher-order interaction) can be associated with a dynamical variable leading to the notion of topological signals. This change of paradigm has lead to novel understanding of topological synchronization~\cite{millan2020explosive,arnaudon2022connecting,calmon2022dirac,ghorbanchian2021higher,nurisso2023unified} and higher-order diffusion dynamics~\cite{torres2020simplicial,ziegler2022balanced,schaub2020random,muhammad2006control} and to novel signal processing~\cite{barbarossa2020topological,schaub2021signal,calmon2023dirac} and topological neural network algorithms \cite{bodnar2021weisfeiler,bodnar2022neural}.
In particular, higher-order diffusion dynamics is among the most basic topological dynamical processes, describing diffusion from $n$-dimensional simplices to $n$-dimensional simplices going either one dimension up or one dimension down. For instance, for $n=1$ higher-order diffusion captures diffusion from edges to edges going either through nodes or through triangles. The main operator driving higher-order diffusion is the Hodge Laplacian \cite{eckmann1944harmonische,horak2013spectra,lim2020hodge} which encodes important information about the simplicial complex topology and is a key player in understanding the interplay between topology and dynamics.

Simplicial complexes have an important limitation as they usually encode only for undirected higher-order interactions. However, in applications, there is an increasing need to include also directional simplicial complexes~\cite{reimann2017cliques,arnaudon2022connecting,courtney2018dense}.
Additionally, from the perspective of algebraic topology, an important challenge is related to the definition of the topology of directed simplicial complexes for which several proposals have been recently suggested~\cite{ruben_directed,grigor2020path,suwayyid2023persistent}.

While treating directionality on simplicial complexes is only in its infancy, on networks considering directional edges is a much more widely explored topic. Among many different approaches, recently it has been shown that the Magnetic Laplacian is a fundamental algebraic operator that captures the directionalities of the edges while preserving a real and positive spectrum.   The Magnetic Laplacian has its roots in theoretical physics and gauge theory \cite{Shubin_Magnetic,Kogut,Kogut_Susskind}, but recently has been shown to provide a valuable tool for node embeddings~\cite{FanuelMichael2018MEft,Post_magnetic,FanuelMichael2017Mefc,gong2021directed,tian2023structural} and for the formulation of new neural network architectures \cite{The_Magnet}. The Magnetic Laplacian is a Hermitian operator that acts on complex-valued variables associated with the nodes and enforces a rotation of the complex phase at every edge. Thus the Magnetic Laplacian can be interpreted as a Laplacian with complex-valued weights on its edges, which is one of the emerging topics in network theory
\cite{tian2023structural,bottcher2022complex,bottcher2024dynamical}. Interestingly, the Magnetic Laplacian can be seen as a generalization of the Connection Laplacian~\cite{singer2012vector, bandeira2013cheeger, chung2013local} acting on a vector field defined on each node of the network and inducing a rotation of the vector in correspondence of each edge.

In this work, we are motivated by the success of the Magnetic Laplacian in treating directed networks and we formulate Higher-order Connection  Laplacians for the study of directed simplicial complexes of dimension two. The $1$-order Connection Laplacians, for example, rotate complex valued 2-dimensional vectors associated with edges when transported across a node or a triangle according to rules depending on the relative directions of the simplicial complex. In particular, we show that in order to take into account all possible configurations of relative directions of simplices we need to make use of the Pauli matrices as well as of an additional rotation of the complex phases.
The higher-order Connection Laplacians are here adopted in order to define higher-order diffusion on directed simplicial complexes that can reveal important dynamical effects induced by the ``frustration" of the relative directions of the simplices as demonstrated from the study of instructive synthetic examples.

By formulating the higher-order Connection Laplacians this work demonstrates that a much unexplored yet very promising research direction in higher-order networks is the investigation of the dynamics of topological signals combined with rich algebraic structures. Other examples of this emerging topic are the adoption of the Dirac operator~\cite{bianconi2021topological} and its associated gamma matrices~\cite{muolo2024three} to study for instance Turing and Dirac-induced patterns of topological signals,  higher-order synchronization dynamics~\cite{lohe2010quantum,lohe2021higher}, the use of sheaves in opinion dynamics~\cite{hansen2021opinion},  and the extensive literature on physics-inspired neural networks~\cite{bodnar2022neural,barbero2022sheaf}.

\section{Combinatorial, Connection and Magnetic Laplacians on graphs}
We consider a graph $G=(V,E)$ with $N_0$ vertices and $N_1$ edges. We are interested in the scenario in which this real graph describes a complex system; in this case, it is also referred to as a network. Here we discuss three definitions of Laplacians: the Combinatorial Graph Laplacian \cite{LuxburgUlrike2007Atos, banerjee2008spectrum, von2008consistency, ChungFanR.K.1997Sgt}, the Connection Laplacian \cite{singer2012vector, bandeira2013cheeger}, and the Magnetic Laplacian \cite{FanuelMichael2018MEft, FanuelMichael2017Mefc}. The Laplacians are algebraic topology operators that can be used to describe diffusion on the graph and have wide applications in network science \cite{FanuelMichael2017Mefc, white2005spectral, spec_hkk}, non-linear dynamics \cite{torres2020simplicial, ziegler2022balanced, lambiotte_schaub_2022}, as well as in signal processing \cite{ghorbanchian2021higher, krishnagopal2021spectral, roddenberry2022hodgelets,  7395367} and machine learning \cite{LuxburgUlrike2007Atos, ng2001spectral, Wang2016}.

\subsection{Combinatorial Graph Laplacian}
The Combinatorial  Graph Laplacian $L^{(0)}$ \cite{LuxburgUlrike2007Atos, ChungFanR.K.1997Sgt} is probably the most widely used Laplacian for undirected graphs and networks.
Let us assume $G=(V,E)$ is simple (unweighted, undirected and without self-loops and reciprocated edges) and has  adjacency matrix $A^{(0)}$; then the Combinatorial Graph Laplacian  $L^{(0)}$ is the  $N_0\times N_0$ matrix  defined as 
\begin{equation}
    L^{(0)} = D^{(0)} - A^{(0)},
\end{equation}
where $D^{(0)}$ is the diagonal matrix having the degrees of the vertices in the network, i.e., $D^{(0)}_{ii}=\sum_{j\in V}^{N_0}A^{(0)}_{ij}$ for all vertices $i$ of the graph.
This Laplacian provides the paradigmatic example of Laplacian operator that reflects the graph's discrete geometry and topology in its spectral properties \cite{ChungFanR.K.1997Sgt} strongly affecting the dynamics unfolding on the graph. Therefore  $L^{(0)}$ provides a key way to relate the structure and dynamics of networks and has a key role in Network Science \cite{lambiotte_schaub_2022} and Machine Learning \cite{he2020graph}.
However, a limitation of the Combinatorial Graph Laplacian is the fact that the original definition only applies to undirected networks.
In the following, we will discuss in detail the Magnetic Laplacian, which addresses this limitation by the adoption of Hermitian matrices.
\subsection{Magnetic Laplacian on Directed Graph}
Consider an unweighted directed graph $G = (V, E)$ with vertex set $V$ and directed edge set $E$ connecting pairs of distinct vertices (no self-loop and no reciprocated edges). Let $a$ indicate the binary directed adjacency matrix, and let us define $A$ as the adjacency matrix of its corresponding unweighted network, i.e. $A_{ij}^{(0)} = 1$ if $(i,j)$ exists (in any direction) and $A_{ij}^{(0)}=0$ otherwise. The Magnetic Laplacian \cite{Shubin_Magnetic,FanuelMichael2018MEft} $L^{(g)}$ is a $N_0\times N_0$ Hermitian matrix  defined as 
\begin{equation}
    L^{(g)} = D^{(0)} - T^{(g)} \circ  A^{(0)}
\end{equation} 
where $\circ$ indicates the Hadamard product and $T^{(g)}_{ij}=\exp(\delta_{ij})$ assigns a rotation in the complex phase to a directed link, $\delta_{ij} = -2\pi g (a_{ij}-a_{ji})$, with $g$ being a constant parameter. Therefore, $\delta_{ij} =  -2\pi g$ if $i \rightarrow j$; $\delta_{ij} =  2\pi g$ if $i \leftarrow j$; and $\delta_{ij} = 0$ otherwise. Additionally, the definition of the Magnetic Laplacian involves the diagonal degree matrix  $D^{(0)} $ is defined as in the previous paragraph.

For convenience, denote $\delta = 2\pi g$. Then 
\begin{equation}
     T^{(g)}_{ij} =
    \begin{cases}
      e^{-\iu \delta}, & \text{if}\ i \rightarrow j  \\
      e^{\iu \delta}, & \text{if}\ j \rightarrow i \\
      1, & \text{otherwise.}
    \end{cases}
\end{equation}

The Magnetic Laplacian acts on complex-valued cochains that can be represented as the vector $\bm\nu\in \mathbb{C}^{N_0}$, having complex values on each of the nodes. The Magnetic Laplacian  is associated with the quadratic form 
\begin{equation}
{\bm \nu}^{\dag}L^{(g)}\bm\nu=\frac{1}{2}\sum_{i,j}A_{ij}\|\nu_i-T^{(g)}_{ij}\nu_j\|^2.
\end{equation}
It is therefore apparent that the Magnetic Laplacian has a real and non-negative spectrum. Its associated quadratic form can be combined with machine learning algorithms to define efficiently network embeddings~\cite{FanuelMichael2018MEft,Post_magnetic,FanuelMichael2017Mefc,gong2021directed,tian2023structural} that can reveal for instance non-trivial cyclic patterns or connections among clusters or communities of the network  and neural networks~ \cite{The_Magnet}. 

\subsection{Connection  Laplacian}
The Connection Laplacian \cite{singer2012vector, bandeira2013cheeger, chung2013local} generalizes the definition of the Magnetic Laplacian and can be used to describe the motion of $d$-dimensional vectors along edges. 
The Connection Laplacian, denoted $L_{0}^{(c)}$, is a $N_0\times N_0$ matrix defined as  
\begin{equation}
L^{(c)}_{0} = D^{(0)}\otimes {I_d} - T^{[0],c} \circ (A^{(0)}\otimes  {\bf 1}_d),
\end{equation} where ${\bf 1}_d$ is the $d\times d$  matrix having all elements equal to one,  and the rotation matrix $T^{[0],C}$ is given by $T^{[0],c}_{ij}=O_{ij}$ where $O_{ij} \in \mathbb{R}^{d \times d}$ is the rotation matrix which represents the orthogonal transformation \(SO(d)\) satisfying \(O_{ij}O_{ji} = I_d\).

The Connection Laplacian acts on real-valued cochains $\bm \nu$ where each element of the cochain associated to a vertex $i$ is a vector  \(\nu_i \in \mathbb{R}^d\). Thus the quadratic form associated with the Connection Laplacian is given by 
\begin{equation}
{\boldsymbol{\nu}^TL^{(c)}\boldsymbol{\nu}}= \frac{1}{2}{\sum_{i,j \in V} A_{ij}\| \nu_i -  O_{ij}\nu_j\|^2}.
\end{equation}
A Connection Laplacian is called {\em consistent} if, for every cycle in the graph, the total rotation is equivalent to the identity matrix. This property ensures that one can always rotate back to the same phase after completing any directed cycle \cite{ChungFan2012RaSa}. Moreover, when a Connection Laplacian is consistent, there exist exactly $d$ eigenvalues with a value of $0$. 

It can be easily shown that the  Magnetic Laplacian \cite{FanuelMichael2018MEft,FanuelMichael2017Mefc} is a special case of Connection Laplacian for directed graphs with a $U(1)$ or $SO(2)$ transformation. 
As we will see in the following, our manuscript proposes to generalize the notion of Graph Connection Laplacian to directed simplicial complexes with the goal of capturing in the spectral properties of these operators the information encoded in the directions of their simplices, in the same spirit as the use of the Magnetic Laplacian to capture the directionalities of edges in graphs and networks.

\section{Simplicial complexes}
Simplicial complexes are a class of higher-order networks that encode higher-order interactions in complex systems.
In this work, we will consider $2$-dimensional simplicial complexes formed by vertices ($0$-simplices), edges ($1$-simplices) and triangles ($2$-simplices). More formally, a  {\em simplex} of dimension $k$ (or $k$-simplex) is formed by $n+1$ vertices, and the {\em face} of a simplex is a simplex of dimension $k'<k$ formed by a proper subset of the vertices of the original simplex.  A  simplicial complex $\mathcal{K}$ is a set of simplices closed under the inclusion of the faces of each of its simplices. The {\em dimension} of a simplicial complex is the largest dimension of its simplices. The use of simplicial complexes~\cite{battiston2020networks,bick2021higher,torres2021and,salnikov2018simplicial} is becoming popular in Network Science and complex systems research as it allows us to model complex systems such as the brain,  protein assemblies, or social interactions, where each simplex characterizes a given higher-order interaction among their elements.

Simplicial complexes allow for higher-order extensions of Combinatorial Graph Laplacians called Hodge Laplacians that can reveal important properties of higher-order diffusion between $n$-simplices to $n$-simplices.
In order to define the Hodge Laplacian it is important to associate to each simplex an orientation, typically chosen to be induced by the node labels. Here and in the following, unless clearly specified, we always adopt this convention as it has the advantage that the spectral properties of the Hodge Laplacians will be independent of the vertex labelling. Note most importantly that the notion of orientation is distinct from the notion of direction. For instance an oriented $1$-dimensional simplex (an oriented graph) will attribute to each edge an orientation and not a direction. The orientation can be used for instance to determine if a flux through an edge is positive or negative;  for instance if the edge $[1,2]$ is oriented positively, a flux going from vertex $1$ to vertex $2$ will be positive, and a flux from vertex $2$ to vertex $1$ will be negative. From this example, it is clear that the orientation of an edge is a different notion with respect to direction, as a directed link will allow only flux to go in one direction.

Hodge Laplacians are receiving increased interest because of their ability to encode the topology of simplicial complexes at the same time as offering the correct algebraic topology operator to define higher-order diffusion and to model, treat and process higher-order topological signals. In particular, the $k$-order Hodge Laplacian acts on $k$ cochains, which can be represented as vectors assigning a dynamical value to each $k$-simplex of the simplicial complex. The action of the $k$-Laplacian on the $k$-cochain can describe diffusion from $k$-simplex to $k$-simplex passing either though $(k-1)$ or to $(k+1)$ simplices.

Here our intention is to give a brief computational introduction to Hodge Laplacians without giving the full background of these important algebraic topology operators. For further information we recommend the following literature, devoted to their spectral properties:   \cite{eckmann1944harmonische,hoff2002latent,lim2020hodge}. 

In order to provide an operational definition of Hodge Laplacians let us introduce some notation.
We consider the finite oriented simplicial complex $\mathcal{K}$ formed by $N_k$ simplices of dimension $k$, each denoted as $\alpha_k^i$ with $1\leq i\leq N_k$.
Two distinct $k$-simplices $\alpha_k^i$ and $\alpha_k^j$ are {\em upper adjacent} if they both are faces of a $(k+1)$-simplex, also known as a co-face. Two simplices that are upper-adjacent and have the same relative orientation with respect to the common co-face are indicated as $\sigma_k^1 \sim_U \sigma_k^2$, while otherwise we indicate $\alpha_k^i \not \sim_U \alpha_k^j$. Two distinct $k$-simplices are {\em lower adjacent} if they share a common face. Two simplices that are lower-adjacent and have the same relative orientation with respect to the common face are indicated as $\sigma_k^1 \sim_L\sigma_k^2$, while otherwise, we indicate $\alpha_k^i \not \sim_L \alpha_k^j$. The $k$-order Hodge Laplacian  is a $N_k\times N_k$ matrix given by 
\begin{equation}
\hL_k=\hL^{up}_k+\hL^{down}_k,
\end{equation}
where  we use the convention $\hL_0^{c, down}=0$ and in all other cases the up and the down Hodge Laplacians $\hL^{up}_k$ and $\hL^{c, down}_k$ have elements
\begin{equation}
    [\hL_k^{up}]_{ij} = \begin{cases}
        \deg_U(\alpha_k^i), & \text{if } i = j\\
        1, & \text{if } \alpha_k^i \sim_U \alpha_k^j\\
        -1, & \text{if } \alpha_k^i \not \sim_U \alpha_k^j\\
        0, & \text{otherwise; } 
    \end{cases},\quad
    [\hL_k^{down}]_{ij} = \begin{cases}
        k+1, & \text{if } i = j\\
        1, & \text{if } \alpha_k^i \sim_L \alpha_k^j\\
        -1, & \text{if } \alpha_k^i \not \sim_L \alpha_k^j\\
        0, & \text{otherwise. } 
    \end{cases}
\end{equation}
From this definition, it follows that as long as $k=0$ the Hodge Laplacian reduces to the Combinatorial Graph Laplacian while for  $k > 0$ the Hodge Laplacian has elements
\begin{equation}
    [\hL_k]_{ij} =  \begin{cases}
        \deg_U(\alpha_k^i) + k+1, & \text{if } i = j\\
        1, & \text{if } \alpha_k^i \sim_L \alpha_k^j \text{ and they are not upper adjacent}\\
        -1, & \text{if } \alpha_k^i \not \sim_L \alpha_k^j \text{ and they are not upper adjacent}\\
        0, & \text{otherwise. } 
    \end{cases}
\end{equation}
Hodge Laplacians encode fundamental topological properties of the simplicial complex. In particular, one of the most celebrated properties of Hodge Laplacians is that the dimension of the kernel of the $n$-Hodge Laplacian is given by the $n$-th Betti number.
Another crucial property of Hodge Laplacians is that they obey Hodge decomposition and thus allow, for instance, the decomposition of edge signals into harmonic, solenoidal, and irrotational components.

\begin{figure}
\centering
\includegraphics[width=0.7\textwidth]{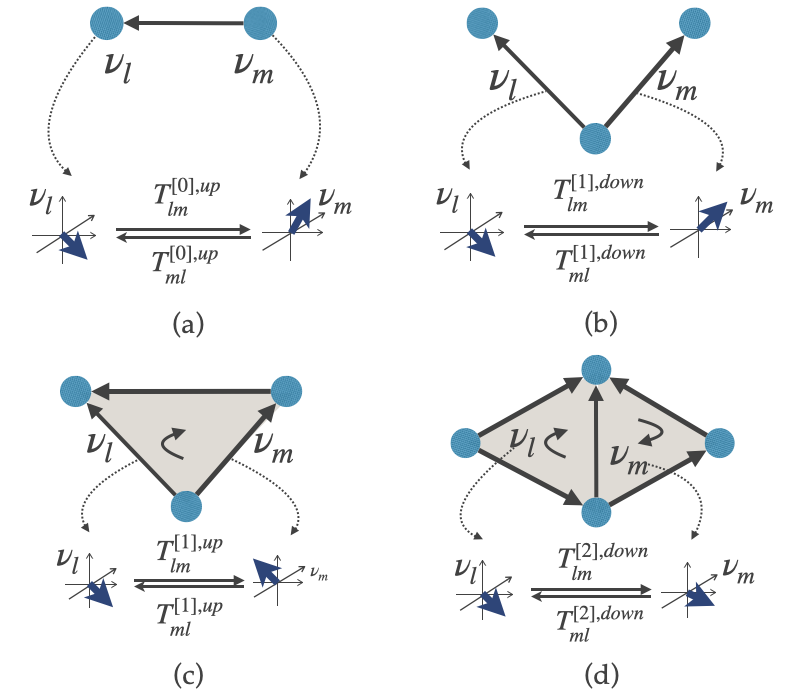}
\caption{Higher-order Connection Laplacians rotate vectors along nodes, edges, or triangles. (a) 0-up Connection Laplacian rotates vectors defined on vertices along edges. (b) 1-down Connection Laplacian rotates vectors defined on edges along nodes. (c) 1-up Connection Laplacian rotates vectors defined on edges along triangles. (d) 2-down Connection Laplacian rotates vectors defined on triangles along edges.}
\label{fig:DirectedHodge}
\end{figure}

\section{Higher-order Connection Laplacians of $2$-dimensional simplicial complexes}

Hodge Laplacians constitute a powerful tool to study the topology of higher-order networks, however, they have the important limitation of only applying to undirected (although oriented) simplicial complexes. This is a drawback for studying real higher-order networks and represents also a significant challenge in mathematics in general and applied topology in particular.

Here our goal is to propose the study of Higher-order Connection Laplacians of $2$-dimensional simplicial complexes, in order to capture spectral properties that are induced by introducing the directionality of the simplices. In doing so we are inspired by the recognized role of the Magnetic Laplacian as the key algebraic topology operator to treat directionality in the graph setting while preserving a real and non-negative spectrum.
In particular, we will define the $k$-order Connection Laplacian that will determine how vectors defined  on simplices are transformed when hopping though upper or lower adjacent simplices (see Figure $\ref{fig:DirectedHodge}$).
As we will see the $0$-Connection Laplacian can be defined trivially as the Combinatorial Graph Laplacian of the underlying network skeleton of the simplicial complex, i.e., the graph formed only by vertices and edges of the simplicial complex. Therefore, this operator can be defined on complex-valued $0$-cochains assigned to each vertex of the graph.
In order to take into account all possible configurations induced by the directions of edges and triangles, the  $1$-Connection Laplacian will however, require us to lift the dimensionality of the $1$-cochain to a $2$-dimensional complex valued vector defined on each edge of the simplicial complex. 
The $2$-Connection Laplacian will be acting on a  $2$-cochain taking as well  a $2$-dimensional complex valued vector  on each triangle of the simplicial complex.  However, another difficulty arises, as to define an {\em Hermitian}  $2$-Connection Laplacian we will have to limit our discussion to simplicial complexes that tessellate $2$-dimensional orientable manifolds.

\begin{table}
\setlength{\tabcolsep}{1mm} 
\def\arraystretch{1.0} 
\centering

\begin{tabular}{|c|c|c|c|}

\hline
\begin{tikzpicture}[>=Stealth]
\draw[fill=black] (0,0) circle (1.5pt);
\draw[fill=black] (3/2,0) circle (1.5pt);
\draw[fill=black] (0.75,1.2) circle (1.5pt);
\node at (-0.2,-0.3) {$j$};
\node at (1.7,-0.3) {$k$};
\node at (0.75,1.5) {$i$};
\draw[thick, ->] (0.75,1.2) -- (0,0);
\draw[thick, ->] (0,0) -- (3/2,0);
\draw[] (3/2,0) -- (0.75,1.2);
\draw[thick, ->] (1,0.43) arc (0:270:0.25);

\end{tikzpicture}

& \begin{tikzpicture}[>=Stealth]
\draw[fill=black] (0,0) circle (1.5pt);
\draw[fill=black] (3/2,0) circle (1.5pt);
\draw[fill=black] (0.75,1.2) circle (1.5pt);
\node at (-0.2,-0.3) {$j$};
\node at (1.7,-0.3) {$k$};
\node at (0.75,1.5) {$i$};
\draw[thick, <-] (0.75,1.2) -- (0,0);
\draw[thick, <-] (0,0) -- (3/2,0);
\draw[] (3/2,0) -- (0.75,1.2);
\draw[thick, <-] (1,0.43) arc (0:270:0.25);

\end{tikzpicture}  

&  \begin{tikzpicture}[>=Stealth]
\draw[fill=black] (0,0) circle (1.5pt);
\draw[fill=black] (3/2,0) circle (1.5pt);
\draw[fill=black] (0.75,1.2) circle (1.5pt);
\node at (-0.2,-0.3) {$j$};
\node at (1.7,-0.3) {$k$};
\node at (0.75,1.5) {$i$};
\draw[thick, <-] (0.75,1.2) -- (0,0);
\draw[thick, <-] (0,0) -- (3/2,0);
\draw[] (3/2,0) -- (0.75,1.2);
\draw[thick, ->] (1,0.43) arc (0:270:0.25);
\end{tikzpicture}

& \begin{tikzpicture}[>=Stealth]
\draw[fill=black] (0,0) circle (1.5pt);
\draw[fill=black] (3/2,0) circle (1.5pt);
\draw[fill=black] (0.75,1.2) circle (1.5pt);
\node at (-0.2,-0.3) {$j$};
\node at (1.7,-0.3) {$k$};
\node at (0.75,1.5) {$i$};
\draw[thick, ->] (0.75,1.2) -- (0,0);
\draw[thick, ->] (0,0) -- (3/2,0);
\draw[] (3/2,0) -- (0.75,1.2);
\draw[thick, <-] (1,0.43) arc (0:270:0.25);
\end{tikzpicture}
\\ (a) & (b) & (c) & (d)
\\ \hline

\begin{tikzpicture}[>=Stealth]
\draw[fill=black] (0,0) circle (1.5pt);
\draw[fill=black] (3/2,0) circle (1.5pt);
\draw[fill=black] (0.75,1.2) circle (1.5pt);
\node at (-0.2,-0.3) {$j$};
\node at (1.7,-0.3) {$k$};
\node at (0.75,1.5) {$i$};
\draw[thick, <-] (0.75,1.2) -- (0,0);
\draw[thick, ->] (0,0) -- (3/2,0);
\draw[] (3/2,0) -- (0.75,1.2);
\draw[thick, ->] (1,0.43) arc (0:270:0.25);
\end{tikzpicture}

& \begin{tikzpicture}[>=Stealth]
\draw[fill=black] (0,0) circle (1.5pt);
\draw[fill=black] (3/2,0) circle (1.5pt);
\draw[fill=black] (0.75,1.2) circle (1.5pt);
\node at (-0.2,-0.3) {$j$};
\node at (1.7,-0.3) {$k$};
\node at (0.75,1.5) {$i$};
\draw[thick, <-] (0.75,1.2) -- (0,0);
\draw[thick, ->] (0,0) -- (3/2,0);
\draw[] (3/2,0) -- (0.75,1.2);
\draw[thick, <-] (1,0.43) arc (0:270:0.25);
\end{tikzpicture}  

&  \begin{tikzpicture}[>=Stealth]
\draw[fill=black] (0,0) circle (1.5pt);
\draw[fill=black] (3/2,0) circle (1.5pt);
\draw[fill=black] (0.75,1.2) circle (1.5pt);
\node at (-0.2,-0.3) {$j$};
\node at (1.7,-0.3) {$k$};
\node at (0.75,1.5) {$i$};
\draw[thick, ->] (0.75,1.2) -- (0,0);
\draw[thick, <-] (0,0) -- (3/2,0);
\draw[] (3/2,0) -- (0.75,1.2);
\draw[thick, ->] (1,0.43) arc (0:270:0.25);
\end{tikzpicture}

& \begin{tikzpicture}[>=Stealth]
\draw[fill=black] (0,0) circle (1.5pt);
\draw[fill=black] (3/2,0) circle (1.5pt);
\draw[fill=black] (0.75,1.2) circle (1.5pt);
\node at (-0.2,-0.3) {$j$};
\node at (1.7,-0.3) {$k$};
\node at (0.75,1.5) {$i$};
\draw[thick, ->] (0.75,1.2) -- (0,0);
\draw[thick, <-] (0,0) -- (3/2,0);
\draw[] (3/2,0) -- (0.75,1.2);
\draw[thick, <-] (1,0.43) arc (0:270:0.25);
\end{tikzpicture}
\\ (e) & (f) & (g) & (h)
\\ \hline
\end{tabular}
\caption{Schematic representation of the eight configurations induced by the directionality of edges and triangles which are distinguished by the Higher-order Connection Laplacian $\hL^{c, up}_{1}$.}
\label{fig:Lm1u}
\end{table}

\subsection{ Higher-order Connection Laplacian for  $2$-dimensional simplicial complexes}
Consider a simplicial complex $\mathcal{K}$ of dimension $2$, including three different types of simplices: nodes, edges, and triangles. We assume that the simplicial complex is unweighted and directed. The undirected version of this simplicial complex is given by the same simplicial complex that is oriented instead of directed, with nodes, edges, and triangles having an orientation induced by the vertex labels, and with nodes having a trivial positive orientation. The directed simplicial complex has both edges and triangles  associated with a direction, either concordant or discordant with their own orientation of its undirected version.
Here our goal is to generalize the notion of the Connection Laplacian to higher-order networks in order to capture the information encoded in the directionality of the simplices and to define a higher-order diffusion that takes into account the directions of the simplices on top of their orientation.
Note however that we do not intend here to define the topology of the directed simplicial complexes, and that the Higher-order Connection Laplacians have significant differences with respect to the Hodge Laplacian of undirected networks. One of the main differences is that in general the Higher-order Connection Laplacian will not obey Hodge decomposition, and in addition, the $1$-up Connection Laplacian and the $1$-down Connection Laplacian might not even commute. 

\subsection{0-up Connection Laplacian}

For the $0$-Connection Laplacian,  by adopting a minimal assumption, we propose to use the Magnetic Laplacian of the directed simplicial complex skeleton formed exclusively by the vertices and the directed edges of the simplicial complex.
The graph Magnetic Laplacian uses complex numbers to encode the direction of the edges and simultaneously builds a Hermitian operator with real eigenvalues that can be useful for machine learning applications.
Note that the vertices of a network have all the same (trivial) direction so that in a network it is sufficient to take into account the two different directions of the edges (accounted for by the sign of the argument of the corresponding complex number).
It follows that the matrix element $L^{(g)}_{ij}$ depends only on the direction of the link $(i,j)$ which can be $i\rightarrow j$ or $i\leftarrow j$.
A question that arises is how to leverage this to define the $1$-order Connection Laplacian or even the $2$-down Connection Laplacian.
\subsection{1-up Connection  Laplacian}
\label{1up}
When one considers the 1-Connection Laplacian, the situation is much richer than at the network level. Indeed the matrix element $L^{[1],up}_{[ij][jk]}$ describing higher-order diffusion between edge $[i,j]$ and edge $[j,k]$ will not only depend on the direction of the triangle $[i\rightarrow j\rightarrow k]$ and  $[k\rightarrow j\rightarrow i]$ but also on the directions of the edges $[i,j]$ and $[j,k]$.
Taking into account all the possible combinations of the directions of the two edges and the upper incident triangle leads to eight possible configurations (see Figure $\ref{fig:Lm1u}$).
If follows that it is not possible to use only the phase of complex numbers to distinguish between these 8 possible configurations.
We therefore consider a complex-valued $1-cochain$ $\bm \nu$ of elements $\nu_{\ell}\in \mathbb{C}^2$ defined on each edge $\ell=[i,j]$ of the simplicial complex.
We then consider a  $1$-Connection Laplacian that will enforce rotations induced by the four Pauli matrices in conjunction with rotation of the complex phases of the elements $\nu_{\ell}$.
Since the Pauli matrices, together with the identity matrix, form a set of four, and the complex phase can rotate in two different directions, we can take into account all eight combinatorial options induced by the direction of the two incident edges and the shared triangle.
This greatly enriches the complexity of the definition of the $1$ Connection Laplacian.
To construct the 1-up  Connection Laplacian, we start by constructing the Bochner matrix $\bm{\mathcal{B}}_1^{up}$ \cite{forman} of the 1-up Hodge Laplacian 
 $\hL_{[1]}^{c, up}$ of the undirected version of the simplicial complex whose elements are given by 
 \begin{equation}
 [{\mathcal{B}}_1^{up}]_{lm}=\begin{cases} [\hL_1^{c, up}]_{lm} &\text{if } l\neq m,\\
 \sum_{n}|[\hL_1^{c, up}]_{ln}|  &\text{if } l\neq m.
 \end{cases}
 \end{equation}
 This Bochner matrix is semi-definite positive and can be written as the difference of a diagonal part $D^{(1),up}$  and a non-diagonal part $A_1^{up}$, i.e. 
\begin{equation}
\mathcal{B}_{[1]}^{up}=D^{(1),up}-A_{[1]}^{up}.
\end{equation}
Here, the diagonal matrix $D^{(1),up}$ has elements given by $[D^{(1),up}]_{ll} = 2\deg_U(\sigma_1^l)$, which are twice the number of triangles incident to edge 
$l$.

Moreover by indicating with $\alpha_1^l, \alpha_1^m$ the $l$-th and the $m$-th oriented edges (1-simplices), we  have
\begin{equation}
    [A_{1}^{up}]_{lm} = \begin{cases}
        0, & \text{if } l = m\\
        -1, & \text{if } \alpha_1^l \sim_U \alpha_1^m\\
        1, & \text{if } \alpha_1^l \not \sim_U \alpha_1^m\\
        0, & \text{otherwise.} 
    \end{cases}
\end{equation}

With these definitions we define the 1-up Connection Laplacian, $\hL_{1}^{c, up}$,  acting on  1-cochains $\bm \nu$ of elements $\nu_l\in \mathbb{C}^2$ associated to the generic edge $\alpha_1^l$ of the directed simplicial complex as:   
\begin{equation}
\hL^{c, up}_{1} = D^{(1),up}\otimes {I_2} - T^{[1],up} \circ (A_{1}^{up}\otimes {\bf 1}_2),
\end{equation} where ${\bf 1}_2$ is the $2\times 2$  matrix having all elements equal to one.  The rotation matrix $T^{[1],up}$ is a $2N_1\times 2N_1$ matrix that enforces a different type of rotation depending on the directions of the two incident edges and the direction of the incident triangle. Specifically, for a given constant $\delta \in [0, 2\pi)$ we define $T^{[1],up}_{(ij),(jk)}$  as: 
\begin{equation}
T^{[1],up}_{(ij),(jk)} =
\begin{cases}
e^{-\iu \delta} {\bm\sigma}_0, & \text{if}\ i \rightarrow j , j \rightarrow k , i\rightarrow j\rightarrow k  \hspace{2.75em} \mbox{(a)}\\
e^{\iu \delta} {\bm\sigma}_0, & \text{if}\ j \rightarrow i ,k\rightarrow j, i\leftarrow j\leftarrow k  \hspace{2.75em} \mbox{(b)}\\
e^{-\iu \delta}{\bm \sigma}_x, & \text{if}\ j \rightarrow i ,k\rightarrow j,  i\rightarrow j\rightarrow k   \hspace{2.75em} (\mbox{c)}\\
e^{\iu \delta} {\bm \sigma}_x, & \text{if}\ i \rightarrow j , j \rightarrow k ,  i\leftarrow j\leftarrow k   \hspace{2.75em} \mbox{(d)}\\
e^{-\iu \delta}{\bm \sigma}_y, & \text{if}\ j \rightarrow i , j \rightarrow k ,  i\rightarrow j\rightarrow k   \hspace{2.75em} \mbox{(e)}\\
e^{\iu \delta}{\bm \sigma}_y, & \text{if}\ j \rightarrow i ,j\rightarrow k,  i\leftarrow j\leftarrow k   \hspace{2.75em} \mbox{(f)}\\
e^{-\iu \delta}{\bm \sigma}_z, & \text{if}\ i \rightarrow j ,k\rightarrow j, i\rightarrow j\rightarrow k   \hspace{2.75em}\mbox{(g)}\\
e^{\iu \delta}{\bm \sigma}_z, & \text{if}\ i \rightarrow j , k \rightarrow j , i\leftarrow j\leftarrow k   \hspace{2.75em} \mbox{(h)}\\
{\bm\sigma_0}, & \text{otherwise},
\end{cases}
\label{eq:T1u}
\end{equation}
where the letters refer to the labeling of the configurations in Table $\ref{fig:Lm1u}$. Here,  $\bm\sigma$ are the Pauli matrices

\begin{equation}
{\bm\sigma}_0=\begin{pmatrix}1&0\\0&1\end{pmatrix},\quad \bm\sigma_x=\begin{pmatrix}0&1\\1&0\end{pmatrix},\quad\bm\sigma_y=\begin{pmatrix}0&-i\\i&0\end{pmatrix}\quad \bm\sigma_z=\begin{pmatrix}1&0\\0&-1\end{pmatrix}.
\end{equation}

From this definition of the 1-up Connection Laplacian it is straightforward to express the
associated quadratic form as  
\begin{equation}
\bnu^\dag \hL^{c, up}_{1} \bnu =\frac{1}{2}\left(\sum_{\alpha_1^l \not\sim_U \alpha_1^m} \| \nu_l -  T^{[1],up}_{lm}\nu_m\|^2 + \sum_{\alpha_1^l \sim_U \alpha_1^m} \| \nu_l +  T^{[1],up}_{lm}\nu_m\|^2 \right).
\end{equation}
Therefore it is apparent that $\hL^{c, up}_{1}$ is Hermitian and semi-definite positive.


\subsection{1-down Connection Laplacian}
Now, let us define the 1-down connection  Laplacian describing diffusion from edges to edges through vertices. To this end, we will follow a procedure similar to that employed in formulating the 1-up Connection Laplacian. However, there is an important difference. Since vertices have only one (trivial) possible direction, the number of configurations to consider is reduced to four instead of eight. These configurations are induced by the relative directions of the two edges, as illustrated in Table \ref{fig:Lm1d}. Another difference is that, in this case, the 1-down Hodge Laplacian of the undirected version of the simplicial complex and its Bochner matrix coincide. Therefore we  first extract the diagonal and the off-diagonal entries of $\hL_{1}^{down}$:
\begin{table}
\setlength{\tabcolsep}{1mm} 
\def\arraystretch{1.0} 
\centering

\begin{tabular}{|c|c|c|c|}

\hline
\begin{tikzpicture}[>=Stealth]
\draw[fill=black] (0,0) circle (1.5pt);
\draw[fill=black] (3/2,0) circle (1.5pt);
\draw[fill=black] (0.75,1.2) circle (1.5pt);
\node at (-0.2,-0.3) {$j$};
\node at (1.7,-0.3) {$k$};
\node at (0.75,1.5) {$i$};
\draw[thick, ->] (0.75,1.2) -- (0,0);
\draw[thick, ->] (0,0) -- (3/2,0);

\end{tikzpicture}

& \begin{tikzpicture}[>=Stealth]
\draw[fill=black] (0,0) circle (1.5pt);
\draw[fill=black] (3/2,0) circle (1.5pt);
\draw[fill=black] (0.75,1.2) circle (1.5pt);
\node at (-0.2,-0.3) {$j$};
\node at (1.7,-0.3) {$k$};
\node at (0.75,1.5) {$i$};
\draw[thick, <-] (0.75,1.2) -- (0,0);
\draw[thick, <-] (0,0) -- (3/2,0);

\end{tikzpicture}  

&  \begin{tikzpicture}[>=Stealth]
\draw[fill=black] (0,0) circle (1.5pt);
\draw[fill=black] (3/2,0) circle (1.5pt);
\draw[fill=black] (0.75,1.2) circle (1.5pt);
\node at (-0.2,-0.3) {$j$};
\node at (1.7,-0.3) {$k$};
\node at (0.75,1.5) {$i$};
\draw[thick, <-] (0.75,1.2) -- (0,0);
\draw[thick, ->] (0,0) -- (3/2,0);
\end{tikzpicture}

& \begin{tikzpicture}[>=Stealth]
\draw[fill=black] (0,0) circle (1.5pt);
\draw[fill=black] (3/2,0) circle (1.5pt);
\draw[fill=black] (0.75,1.2) circle (1.5pt);
\node at (-0.2,-0.3) {$j$};
\node at (1.7,-0.3) {$k$};
\node at (0.75,1.5) {$i$};
\draw[thick, ->] (0.75,1.2) -- (0,0);
\draw[thick, <-] (0,0) -- (3/2,0);
\end{tikzpicture}
\\ (a) & (b) & (c) & (d)
\\ \hline
\end{tabular}
\caption{Schematic representation of the four configurations induced by the directionality of the edges incident to the same nodes, which are captured by the Higher-order Connection Laplacian $ L^{c, down}_{1}$.}
\label{fig:Lm1d}
\end{table}

\begin{equation}
\hL_{1}^{c, down}= D_{1}^{down} - A_{1}^{down},
\end{equation}
where
\begin{equation}
    [A_{1}^{down}]_{lm} = \begin{cases}
        -1, & \text{if } \alpha_1^l \sim_L \alpha_1^m\\
        1, & \text{if } \alpha_1^l \not \sim_L \alpha_1^m\\
        0, & \text{otherwise, } 
    \end{cases}
\end{equation}
and  $D_{1}^{down}$ as   the diagonal matrix of elements $[D_{1}^{down}]_{ll} = \sum_{m = 1}^n \lvert [A_{1}^{down}]_{lm} \rvert.$
Finally, the $1$-down Connection Laplacian  $\hL_{1}^{c, down}$ is defined by incorporating  information about edge orientation of the undirected version of the simplicial complex with $A_{1}^{down}$ and the edge directions of the actual directed simplicial complex using a rotation matrix $T^{[1],down}$:
\begin{equation}
\hL^{c, down}_{1} = D_{1}^{down}\otimes {I_2} - T^{[1],down} \circ  (A_{1}^{down}\otimes {\bf 1}_2),
\end{equation} 
where for a constant $\delta \in [0, 2\pi)$ the 1-down rotation matrix $T^{[1],down}$ is given by
\begin{equation}
T^{[1],down}_{(ij),(jk)} =
\begin{cases}
e^{\iu \delta} {\bm\sigma_0} , & \text{if}\ i \rightarrow j , j \rightarrow k   \hspace{2em} \mbox{(a)}\\
e^{-\iu \delta} {\bm\sigma_0}, & \text{if}\ j \rightarrow i ,k\rightarrow j   \hspace{2em} \mbox{(b)}\\
{\bm \sigma}_y, & \text{if}\ j \rightarrow i , j \rightarrow k  \hspace{2em} \mbox{(c)}\\
{\bm \sigma}_z, & \text{if}\ i \rightarrow j, k \rightarrow j  \hspace{2em} \mbox{(d)}\\
{\bm\sigma_0}, & \text{otherwise}.
\end{cases}
\end{equation}
Note that the letters in the above equation refer to the label of the configuration in Table $\ref{fig:Lm1d}$. 

From this definition it can be easily shown that  $\hL^{c, down}_{1}$ is Hermitian as both $D_{1}^{down}$ and $A_{1}^{down}$ are real and symmetric; and $T^{[1],down}$ is Hermitian.

Assigning a complex vector $\nu_l \in \mathbb{C}^2$ to edge $l$, and let $\bnu = \{ \nu_1, ..., \nu_n \}$, we have:
\begin{equation}
\bnu^{\dag} \hL^{c, down}_{1} \bnu =\frac{1}{2}\left(\sum_{\alpha_1^l \not\sim_L \alpha_1^m} \lVert  \nu_l -  T^{[1],down}_{lm}\nu_m \rVert^2 + \sum_{\alpha_1^l \sim_L \alpha_1^m} \| \nu_l + T^{[1],down}_{lm}\nu_m\|^2 \right).
\end{equation}
Therefore we can conclude that  $\hL^{c, down}_{1}$ is positive semi-definite.

\subsection{2-down connection  Laplacian for 2-dimensional orientable manifolds}

The extension to Higher-order Connection Laplacians comes with additional challenges.  For instance, in order to extend the above definition to the $2$-down Connection Laplacian, relying solely on the definition of orientation and direction of triangles induced by the node labels is not sufficient to obtain Hermitian matrices. This implies that we cannot extend the above definition to treat the $2$-down diffusion on a general simplicial complex. What we can do, however, is to consider simplicial complexes forming discrete orientable surfaces. For triangles, we can adopt the notion of orientation (and direction) induced by the orientation of the manifold. As for orientable continuous manifolds, for discrete orientable manifolds, we can distinguish two sides of a surface, with the normal vector to the manifold pointing either to one side or the other side of the manifold. The orientation of the manifold is determined by the choice of the normal vector direction. When considering $i, j, k \in V$ such that $[i, j, k]$ or any of its permutations form a 2-simplex, we represent its orientation as $\Delta_{ijk} = 1$ if, by following the right-hand rule and moving along the flow $i\to j\to k$ we obtain a vector pointing in the same direction as the normal vector to the manifold, otherwise  we assign to the $2$ simplex the orientation $\Delta_{ijk} = -1$.

In this way, it can be easily checked that the number of configurations induced by the relative directions of two triangles and their common edges is always eight as in the case in Sec. \ref{1up}. These configurations are shown schematically in Figure $\ref{fig:Lm2d}$. 
As for  the 1-down Laplacian, also the 2-down Laplacian and its the Bochner matrices coincide. Therefore we  first we extract the diagonal and the off-diagonal entries of $\hL_{2}^{c, down}$:
\begin{equation}
\hL_{2}^{c, down}= D_{2}^{down} - A_{2}^{down},
\end{equation}
where
\begin{equation}
    [A_{2}^{down}]_{lm} = \begin{cases}
        -1, & \text{if } \alpha_2^l \sim_L \alpha_2^m\\
        1, & \text{if } \alpha_2^l \not \sim_L \alpha_2^m\\
        0, & \text{otherwise. } 
    \end{cases}
\end{equation}
and  $D_{2}^{down}$ as the diagonal matrix of non zero elements
$    [D_{2}^{down}]_{ll} = \sum_{m = 1}^n \lvert [A_{2}^{down}]_{lm} \rvert.$
Finally, the $2$-down Connection Laplacian  $\hL_{1}^{c, down}$ is defined by incorporating both the triangle directions $A_{2}^{down}$ and the edge directions using a rotation matrix $T^{[2],down}$:
\begin{equation}
\hL^{c, down}_{2} = D_{2}^{down}\otimes {I_2} - T^{[2],down} \circ  (A_{2}^{down}\otimes {\bf 1}_2),
\end{equation} 
where, considering rotations induced by the four Pauli matrices and the complex phase, we have 
\begin{equation}
     T^{[2],down}_{(ijk),(jkk')} =\left\{
    \begin{array}{lllll}
       e^{-\iu \delta} {\bm\sigma_0} , & \text{if  } \Delta_{ijk} = 1 , & \Delta_{kjk'} = -1 , & j \rightarrow k  & \mbox{(a)}\\
     e^{\iu \delta} {\bm\sigma_0} , & \text{if } \Delta_{ijk} = -1 , & \Delta_{kjk'} = 1 , & k \rightarrow j & \mbox{(b)}\\
      e^{-\iu \delta}{\bm \sigma}_x, & \text{if  } \Delta_{ijk} = -1 , & \Delta_{kjk'} = 1 , & j \rightarrow k & \mbox{(c)}\\
      e^{\iu \delta}{\bm \sigma}_x, & \text{if  }\Delta_{ijk} = 1 , & \Delta_{kjk'} = -1 , & k \rightarrow j & \mbox{(d)}\\
      e^{-\iu \delta}{\bm \sigma}_y, & \text{if  } \Delta_{ijk} = 1 , & \Delta_{kjk'} = 1 , & j \rightarrow k & \mbox{(e)}\\
      e^{\iu \delta}{\bm \sigma}_y, & \text{if  } \Delta_{ijk} = 1 , & \Delta_{kjk'} = 1 , & k \rightarrow j & \mbox{(f)}\\
      e^{-\iu \delta}{\bm \sigma}_z, & \text{if } \Delta_{ijk} = -1 , & \Delta_{kjk'} =-1 , & j \rightarrow k & \mbox{(g)}\\
      e^{\iu \delta}{\bm \sigma}_z, & \text{if  }\Delta_{ijk} = -1 , & \Delta_{kjk'} =-1 , & k \rightarrow j & \mbox{(h)}\\
       {\bm\sigma}_0, & \text{otherwise}, & & &
    \end{array}\right.
\end{equation}
with the letters referring to the labels of the configurations in Table \ref{fig:Lm2d}. 
The associated quadratic form reads 
\begin{equation}
\bnu^{\dag} \hL^{c, down}_{2} \bnu =\frac{1}{2}\left(\sum_{\alpha_2^l \not\sim_L \alpha_2^m} \lVert  \nu_l -  T^{[2],down}_{lm}\nu_m \rVert^2 + \sum_{\alpha_2^l \sim_L \alpha_2^m} \| \nu_l + T^{[2],down}_{lm}\nu_m\|^2 \right).
\end{equation}

\begin{table}
\setlength{\tabcolsep}{1mm} 
\def\arraystretch{1.0} 
\centering

\begin{tabular}{|c|c|c|c|}

\hline
\begin{tikzpicture}[>=Stealth]
\draw[fill=black] (0,0) circle (1.5pt);
\draw[fill=black] (3/2,0) circle (1.5pt);
\draw[fill=black] (0.75,1.2) circle (1.5pt);
\draw[fill=black] (0.75,-1.2) circle (1.5pt);
\node at (-0.2,-0.3) {$j$};
\node at (1.7,-0.3) {$k$};
\node at (0.75,1.5) {$i$};
\node at (0.75,-1.5) {$k'$};
\draw[thick, ->] (0,0) -- (3/2,0);
\draw[] (0.75,1.2) -- (0,0);
\draw[] (3/2,0) -- (0.75,1.2);
\draw[] (3/2,0) -- (0.75,-1.2);
\draw[] (0,0) -- (0.75,-1.2);
\draw[thick, ->] (0.91,0.62) arc (45:315:0.23);
\draw[thick, <-] (0.91,-0.3) arc (45:315:0.23);
\end{tikzpicture}

& \begin{tikzpicture}[>=Stealth]
\draw[fill=black] (0,0) circle (1.5pt);
\draw[fill=black] (3/2,0) circle (1.5pt);
\draw[fill=black] (0.75,1.2) circle (1.5pt);
\draw[fill=black] (0.75,-1.2) circle (1.5pt);
\node at (-0.2,-0.3) {$j$};
\node at (1.7,-0.3) {$k$};
\node at (0.75,1.5) {$i$};
\node at (0.75,-1.5) {$k'$};
\draw[thick, <-] (0,0) -- (3/2,0);
\draw[] (0.75,1.2) -- (0,0);
\draw[] (3/2,0) -- (0.75,1.2);
\draw[] (3/2,0) -- (0.75,-1.2);
\draw[] (0,0) -- (0.75,-1.2);
\draw[thick, <-] (0.91,0.62) arc (45:315:0.23);
\draw[thick, ->] (0.91,-0.3) arc (45:315:0.23);
\end{tikzpicture}  

&  \begin{tikzpicture}[>=Stealth]
\draw[fill=black] (0,0) circle (1.5pt);
\draw[fill=black] (3/2,0) circle (1.5pt);
\draw[fill=black] (0.75,1.2) circle (1.5pt);
\draw[fill=black] (0.75,-1.2) circle (1.5pt);
\node at (-0.2,-0.3) {$j$};
\node at (1.7,-0.3) {$k$};
\node at (0.75,1.5) {$i$};
\node at (0.75,-1.5) {$k'$};
\draw[thick, ->] (0,0) -- (3/2,0);
\draw[] (0.75,1.2) -- (0,0);
\draw[] (3/2,0) -- (0.75,1.2);
\draw[] (3/2,0) -- (0.75,-1.2);
\draw[] (0,0) -- (0.75,-1.2);
\draw[thick, <-] (0.91,0.62) arc (45:315:0.23);
\draw[thick, ->] (0.91,-0.3) arc (45:315:0.23);
\end{tikzpicture}

& \begin{tikzpicture}[>=Stealth]
\draw[fill=black] (0,0) circle (1.5pt);
\draw[fill=black] (3/2,0) circle (1.5pt);
\draw[fill=black] (0.75,1.2) circle (1.5pt);
\draw[fill=black] (0.75,-1.2) circle (1.5pt);
\node at (-0.2,-0.3) {$j$};
\node at (1.7,-0.3) {$k$};
\node at (0.75,1.5) {$i$};
\node at (0.75,-1.5) {$k'$};
\draw[thick, <-] (0,0) -- (3/2,0);
\draw[] (0.75,1.2) -- (0,0);
\draw[] (3/2,0) -- (0.75,1.2);
\draw[] (3/2,0) -- (0.75,-1.2);
\draw[] (0,0) -- (0.75,-1.2);
\draw[thick, ->] (0.91,0.62) arc (45:315:0.23);
\draw[thick, <-] (0.91,-0.3) arc (45:315:0.23);
\end{tikzpicture}
\\ (a) & (b) & (c) & (d)
\\ \hline
\begin{tikzpicture}[>=Stealth]
\draw[fill=black] (0,0) circle (1.5pt);
\draw[fill=black] (3/2,0) circle (1.5pt);
\draw[fill=black] (0.75,1.2) circle (1.5pt);
\draw[fill=black] (0.75,-1.2) circle (1.5pt);
\node at (-0.2,-0.3) {$j$};
\node at (1.7,-0.3) {$k$};
\node at (0.75,1.5) {$i$};
\node at (0.75,-1.5) {$k'$};
\draw[thick, ->] (0,0) -- (3/2,0);
\draw[] (0.75,1.2) -- (0,0);
\draw[] (3/2,0) -- (0.75,1.2);
\draw[] (3/2,0) -- (0.75,-1.2);
\draw[] (0,0) -- (0.75,-1.2);
\draw[thick, ->] (0.91,0.62) arc (45:315:0.23);
\draw[thick, ->] (0.91,-0.3) arc (45:315:0.23);
\end{tikzpicture}

& \begin{tikzpicture}[>=Stealth]
\draw[fill=black] (0,0) circle (1.5pt);
\draw[fill=black] (3/2,0) circle (1.5pt);
\draw[fill=black] (0.75,1.2) circle (1.5pt);
\draw[fill=black] (0.75,-1.2) circle (1.5pt);
\node at (-0.2,-0.3) {$j$};
\node at (1.7,-0.3) {$k$};
\node at (0.75,1.5) {$i$};
\node at (0.75,-1.5) {$k'$};
\draw[thick, <-] (0,0) -- (3/2,0);
\draw[] (0.75,1.2) -- (0,0);
\draw[] (3/2,0) -- (0.75,1.2);
\draw[] (3/2,0) -- (0.75,-1.2);
\draw[] (0,0) -- (0.75,-1.2);
\draw[thick, ->] (0.91,0.62) arc (45:315:0.23);
\draw[thick, ->] (0.91,-0.3) arc (45:315:0.23);
\end{tikzpicture}  

&  \begin{tikzpicture}[>=Stealth]
\draw[fill=black] (0,0) circle (1.5pt);
\draw[fill=black] (3/2,0) circle (1.5pt);
\draw[fill=black] (0.75,1.2) circle (1.5pt);
\draw[fill=black] (0.75,-1.2) circle (1.5pt);
\node at (-0.2,-0.3) {$j$};
\node at (1.7,-0.3) {$k$};
\node at (0.75,1.5) {$i$};
\node at (0.75,-1.5) {$k'$};
\draw[thick, ->] (0,0) -- (3/2,0);
\draw[] (0.75,1.2) -- (0,0);
\draw[] (3/2,0) -- (0.75,1.2);
\draw[] (3/2,0) -- (0.75,-1.2);
\draw[] (0,0) -- (0.75,-1.2);
\draw[thick, <-] (0.91,0.62) arc (45:315:0.23);
\draw[thick, <-] (0.91,-0.3) arc (45:315:0.23);
\end{tikzpicture}

& \begin{tikzpicture}[>=Stealth]
\draw[fill=black] (0,0) circle (1.5pt);
\draw[fill=black] (3/2,0) circle (1.5pt);
\draw[fill=black] (0.75,1.2) circle (1.5pt);
\draw[fill=black] (0.75,-1.2) circle (1.5pt);
\node at (-0.2,-0.3) {$j$};
\node at (1.7,-0.3) {$k$};
\node at (0.75,1.5) {$i$};
\node at (0.75,-1.5) {$k'$};
\draw[thick, <-] (0,0) -- (3/2,0);
\draw[] (0.75,1.2) -- (0,0);
\draw[] (3/2,0) -- (0.75,1.2);
\draw[] (3/2,0) -- (0.75,-1.2);
\draw[] (0,0) -- (0.75,-1.2);
\draw[thick, <-] (0.91,0.62) arc (45:315:0.23);
\draw[thick, <-] (0.91,-0.3) arc (45:315:0.23);
\end{tikzpicture}
\\ (e) & (f) & (g) & (h)
\\ \hline
\end{tabular}
\caption{Schematic representation of the eight configurations induced by the directionality of two lower adjacent triangles and their common lower adjacent edge on $2$-dimensional orientable manifolds. These configurations are  captured by the Higher-order Connection Laplacian  $ L^{c,down}_{[2]}$.}
\label{fig:Lm2d}
\end{table}

Note that the $2$-down  Connection Laplacian can be also extended to define the $n$-down Connection Laplacian of $n$-dimensional orientable manifolds, as the definition of this Connection Laplacian will always entail treating eight different configurations of relative directions of the two $n$-dimensional simplices and their incident $(n-1)$-dimensional simplex.

As we have seen, our formalism has allowed us to  define the  $0$-up, the $1$-up, the $1$-down, and the $2$-down Connection Laplacians of orientable, two-dimensional discrete manifolds. These operators allow us to  distinguish between all possible configurations of the directions of the simplices by inducing rotations  enforced through the matrices  $T^{[n],up/down}$ leveraging on the use of the Pauli matrices, and a rotation of the phase in the complex plane.

\section{Higher-order diffusion on directed simplicial complexes }

The Connection Laplacians defined in the previous section can be used to define higher-order diffusion processes that will reflect the directionality of the simplicial complex extending previous work on higher-order diffusion over undirected simplicial complexes~\cite{torres2020simplicial,ziegler2022balanced,schaub2020random,muhammad2006control}.
If we focus on the diffusion induced by the $1$-Connection Laplacian we can define three types of dynamical processes describing diffusion from edge to edge going exclusively through triangles (upper diffusion), going exclusively through nodes (lower diffusion), or going either through triangles or  nodes (combined diffusion). These three types of diffusion process describe the evolution of the $1-cochain$ $\bnu$ whose elements $\nu_l$ defined on each edge $l$ are complex valued $2$-dimensional vectors, i.e. $\nu_l\in \mathbb{C}^2$.
Specifically, the upper, lower, and combined diffusion are defined by the following dynamical processes for $\bnu=\bnu(t)$:
\bea
     \frac{d \bnu(t)}{dt} = -\hL_{1}^{c, up} \bnu(t),\nonumber \\
     \frac{d \bnu(t)}{dt} = -\hL_{1}^{c, down} \bnu(t),\nonumber \\
     \frac{d \bnu(t)}{dt} = -(\hL_{1}^{c, up}+ \hL_{1}^{c, down}) \bnu(t).
\eea
Note that since in general $\hL_{1}^{c, up}$ and $\hL_{1}^{c, down}$ do not obey Hodge decomposition, the combined diffusion process cannot be easily recast into the upper and the lower diffusion dynamics as happens in diffusion determined by the Hodge Laplacian in undirected simplicial complexes \cite{torres2020simplicial,taylor2015topological}.

The properties of the higher-order diffusion (of order $1$) on directed simplicial complexes will depend strongly on the structure of the simplicial complex and the directionality of the edges. Here we provide a discussion of these higher-order diffusion processes in different cases of directed triangles and on a directed torus.
For each case considered, we will investigate the spectral properties of the $1$-up and $1$-down Connection Laplacians as a function of the parameter $\delta$ and we will computer the commutator
\begin{equation}
    [\hL_{1}^{c, up}, \hL_{1}^{c, down}] = \hL_{1}^{c, up}\hL_{1}^{c, down}-\hL_{1}^{c, down}\hL_{1}^{c, up}.
\end{equation}

\subsection{Directed Simplicial Triangles}
\label{chp3:triangle}

Let us now explore some examples of directed triangles formed by three vertices $[1], [2], [3]$, three edges $[1,2]$, $[1,3]$, $[2,3]$, and one triangle $[1,2,3]$, where the edges and the triangles have a given direction, leading to the four distinct  scenarios depicted in Table \ref{fig:triangle}. 
In addition to analysing of the spectral properties of these simplicial triangles, for each considered case, we will also  analyse  the quadratic form of the Higher-order Connection Laplacians and determine conditions under which it can be minimized to zero. Finally,  we will provide a numerical investigation of the dynamical properties of the induced higher-order diffusion processes.

\begin{table}
\setlength{\tabcolsep}{1mm} 
\def\arraystretch{1.0} 
\centering

\begin{tabular}{|c|c|c|c|}

\hline
\begin{tikzpicture}[>=Stealth]
\draw[fill=black] (0,0) circle (1.5pt);
\draw[fill=black] (3/2,0) circle (1.5pt);
\draw[fill=black] (0.75,1.2) circle (1.5pt);
\node at (-0.2,-0.3) {$2$};
\node at (1.7,-0.3) {$3$};
\node at (0.75,1.5) {$1$};
\draw[thick, ->] (0.75,1.2) -- (0,0);
\draw[thick, ->] (0,0) -- (3/2,0);
\draw[thick, ->] (3/2,0) -- (0.75,1.2);
\draw[thick, ->] (1,0.43) arc (0:270:0.25);

\end{tikzpicture}

& \begin{tikzpicture}[>=Stealth]
\draw[fill=black] (0,0) circle (1.5pt);
\draw[fill=black] (3/2,0) circle (1.5pt);
\draw[fill=black] (0.75,1.2) circle (1.5pt);
\node at (-0.2,-0.3) {$2$};
\node at (1.7,-0.3) {$3$};
\node at (0.75,1.5) {$1$};
\draw[thick, ->] (0.75,1.2) -- (0,0);
\draw[thick, ->] (0,0) -- (3/2,0);
\draw[thick, ->] (3/2,0) -- (0.75,1.2);
\draw[thick, <-] (1,0.43) arc (0:270:0.25);

\end{tikzpicture}  

&  \begin{tikzpicture}[>=Stealth]
\draw[fill=black] (0,0) circle (1.5pt);
\draw[fill=black] (3/2,0) circle (1.5pt);
\draw[fill=black] (0.75,1.2) circle (1.5pt);
\node at (-0.2,-0.3) {$2$};
\node at (1.7,-0.3) {$3$};
\node at (0.75,1.5) {$1$};
\draw[thick, ->] (0.75,1.2) -- (0,0);
\draw[thick, ->] (0,0) -- (3/2,0);
\draw[thick, <-] (3/2,0) -- (0.75,1.2);
\draw[thick, ->] (1,0.43) arc (0:270:0.25);
\end{tikzpicture}

& \begin{tikzpicture}[>=Stealth]
\draw[fill=black] (0,0) circle (1.5pt);
\draw[fill=black] (3/2,0) circle (1.5pt);
\draw[fill=black] (0.75,1.2) circle (1.5pt);
\node at (-0.2,-0.3) {$2$};
\node at (1.7,-0.3) {$3$};
\node at (0.75,1.5) {$1$};
\draw[thick, ->] (0.75,1.2) -- (0,0);
\draw[thick, ->] (0,0) -- (3/2,0);
\draw[thick, <-] (3/2,0) -- (0.75,1.2);
\draw[thick, <-] (1,0.43) arc (0:270:0.25);
\end{tikzpicture}
\\  Case 1 & Case 2   & Case 3  & Case 4
\\ \hline
\end{tabular}
\caption{Schematic representation of the 4 distinct types of $2$-dimensional directed simplicial complexes formed by one directed triangles, three directed edges and three vertices.}
\label{fig:triangle}
\end{table}

\subsubsection{Case 1}
\label{chp3:triangle1}
We consider the directed $2$-simplicial complex denoted as Case 1 in Table \ref{fig:triangle}, having edge directions given by $1\to2$, $2\to 3$, and $3\to 1$; and triangle direction $1\rightarrow 2\rightarrow 3$. This represents the scenario where all edge directions conform to the direction of the triangle. Hence, if we traverse from one edge to another, we either align with both the directions of the edge and the triangle, as shown in the first row of (\ref{eq:T1u}); or we go against both the edge and triangle directions, corresponding to the second row in (\ref{eq:T1u}).
\begin{figure}
\centering
\includegraphics[width=0.9\textwidth]{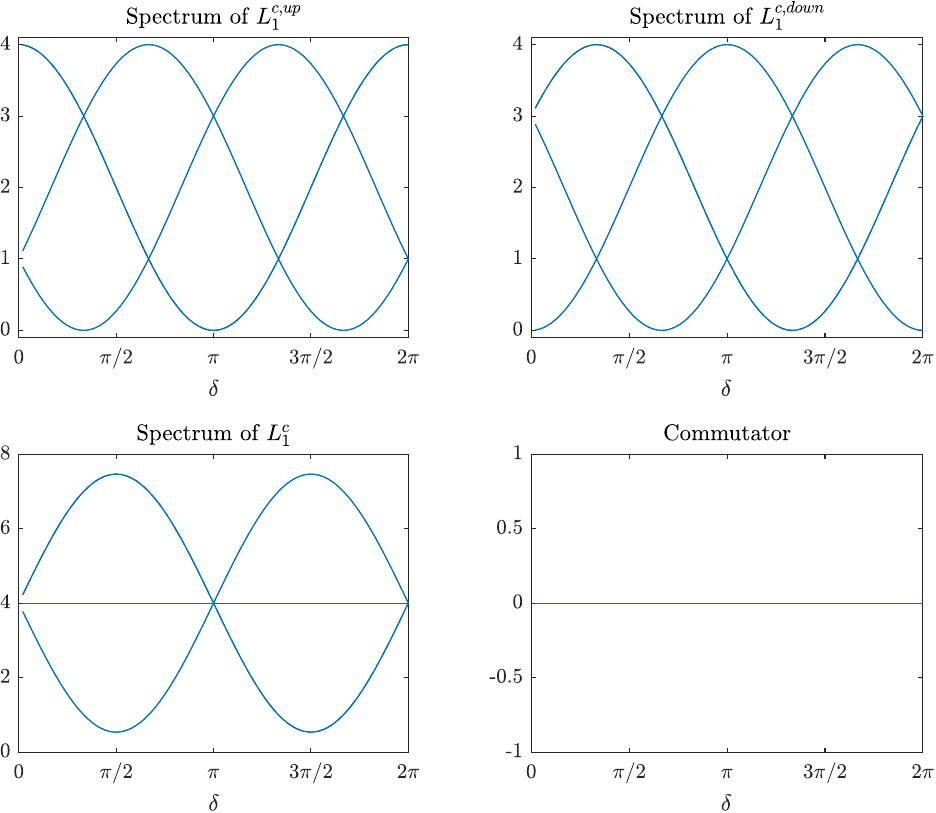}
\caption{The complete spectrum of $\hL_{1}^{c, up}$ (top-left), $\hL_{1}^{c, down}$ (top-right), and $\hL_{1}^{c}$ (bottom-left), and commutator $[\hL_{1}^{c, up}, \hL_{1}^{c, down}] $ (bottom-right) for Case 1 directed simplicial triangles is plotted as a function of $\delta$. }
\label{fig:case1}
\end{figure}
For the $1$-up Connection Laplacian, we obtain the following:
\begin{equation}
\hL_{1}^{c, up}=\bordermatrix{
&[1,2]&[1,3]&[2,3]\cr
[1,2]&2{\bm\sigma_0} &-{\bm\sigma_0}  e^{\iu \delta}&{\bm\sigma_0} e^{-\iu\delta}\cr
[1,3]&-{\bm\sigma_0} e^{-\iu\delta}&2{\bm\sigma_0} &-{\bm\sigma_0} e^{\iu\delta}\cr
[2,3]&{\bm\sigma_0} e^{\iu\delta}&-{\bm\sigma_0} e^{-\iu\delta}&2{\bm\sigma_0} \cr}.
\end{equation}
In the top left panel of Figure (\ref{fig:case1}), we display the eigenvalues of $\hL_{1}^{c, up}$ in relation to $\delta$. The corresponding eigenvalues are:
\begin{equation}
\{ 2 + 2\cos(\delta), 2 + 2 \cos(\delta - 2\pi/3), 2 + 2 \cos(\delta - 4\pi/3)\}.
\end{equation}
To compute the quadratic form, we denote the components of the $1$-cochain as $\bnu$ $\bnu = (\nu_1, \nu_2, \nu_3)' \in \mathbb{C}^6$  corresponding to the three edges $[1,2]$, $[1,3]$, and $[2,3]$ respectively. Here and in the following  we parameterize each component $\nu_l$ of the $1$-cochain  with three angles ($\theta_l$,$\phi_l$ and $\psi_l$) by setting 
\begin{equation}
\nu_l = (\cos(\psi_l)e^{\iu \theta_l}, \sin(\psi_l)e^{\iu \phi_l})' \in \mathbb{C}^2,
\end{equation} 
 where $\psi_l \in [0, \frac{\pi}{2}]$, and $\theta_l, \phi_l \in [0, 2\pi)$. 
Therefore the $1$-up Connection Laplacian is associated to the  quadratic form 
\begin{align}
\bnu^H \hL_{1}^{c, up} \bnu &=  \| \nu_1 - {\bm\sigma_0} e^{\iu\delta} \nu_2 \|^2 + \| \nu_1 + {\bm\sigma_0} e^{-\iu\delta} \nu_3 \|^2 + \| \nu_2 - {\bm\sigma_0} e^{\iu\delta} \nu_3 \|^2.
\end{align}
The quadratic form above equals zero when $\psi_1 = \psi_2 = \psi_3$,  and 
\begin{align}
& \theta_1  \equiv \theta_2 + \delta ,\text{ } \phi_1  \equiv \phi_2 + \delta ,\\ \label{eq:case1_Lu=0_row1}
& \theta_1  \equiv \theta_3 - \delta + \pi ,\text{ } \phi_1  \equiv \phi_3 - \delta + \pi ,\\
& \theta_2  \equiv \theta_3 + \delta , \text{ } \phi_2  \equiv \phi_3 + \delta .
\label{eq:case1_Lu=0_row3}
\end{align} A solution exists only when $\delta = \pi/3, \pi$, or $5\pi/3$, as shown in the top-left panel of Figure \ref{fig:case1}. When $\delta$ takes the aforementioned values, $\bnu$, which satisfy the above conditions, becomes an eigenvector of $\hL_{1}^{c, up}$ associated with an eigenvalue of zero. 

For the $1$-down Connection Laplacian, we have
\begin{equation}
\hL_{1}^{c, down}=\bordermatrix{
&[1,2]&[1,3]&[2,3]\cr
[1,2]&2{\bm\sigma_0} &{\bm\sigma_0}  e^{-\iu \delta}&-{\bm\sigma_0} e^{\iu\delta}\cr
[1,3]&{\bm\sigma_0} e^{\iu\delta}&2{\bm\sigma_0} &{\bm\sigma_0} e^{-\iu\delta}\cr
[2,3]&-{\bm\sigma_0} e^{-\iu\delta}&{\bm\sigma_0} e^{\iu\delta}&2{\bm\sigma_0}\cr }.
\end{equation}
Eigenvalues of $\hL_{1}^{c, down}$, as shown in the top-right plot in Figure \ref{fig:case1}, are: 
\begin{equation}
\{ 2 - 2\cos(\delta + \pi/3), 2 - 2 \cos(\delta), 2 - 2 \cos(\delta - \pi/3)\}.
\end{equation}
The corresponding quadratic form of the $1$-down Connection Laplacians are 
\begin{align}
\bnu^H \hL_{1}^{c, down} \bnu &=  \| \nu_1 + {\bm\sigma_0} e^{-\iu\delta} \nu_2 \|^2 + \| \nu_1 - {\bm\sigma_0} e^{\iu\delta} \nu_3 \|^2 + \| \nu_2 + {\bm\sigma_0} e^{-\iu\delta} \nu_3 \|^2.
\end{align}

It becomes zero when $\psi_1 = \psi_2 = \psi_3$, and
\begin{align}
\label{eq:case1_Ld=0_row1}
&\theta_1 \equiv \theta_2-\delta + \pi \text{, } \phi_1 \equiv \phi_2-\delta + \pi,\\
&\theta_1 \equiv \theta_3-\delta + \pi \text{, } \phi_1 \equiv  \phi_3+\delta,\\
&\theta_2 \equiv \theta_3-\delta + \pi \text{, } \phi_2 \equiv  \phi_3-\delta + \pi.
\label{eq:case1_Ld=0_row3}
\end{align} A solution exists only when $\delta = 0, 2\pi/3$, or $4\pi/3$, as shown in the top-right panel in Figure \ref{fig:case1}. 

Upon combining the $1$-up and $1$-down Connection Laplacians, we obtain the eigenvalues of their sum, as depicted in the bottom-left panel of Figure \ref{fig:case1}. The corresponding eigenvalues are 
\begin{equation}
    \{ 3, 3 + 2\sqrt{3}\sin(\delta), 3 - 2\sqrt{3}\sin(\delta) \}.
\end{equation}
\begin{figure}
    \centering
    \begin{subfigure}[b]{\textwidth}
    \centering
    \includegraphics[width=\textwidth]{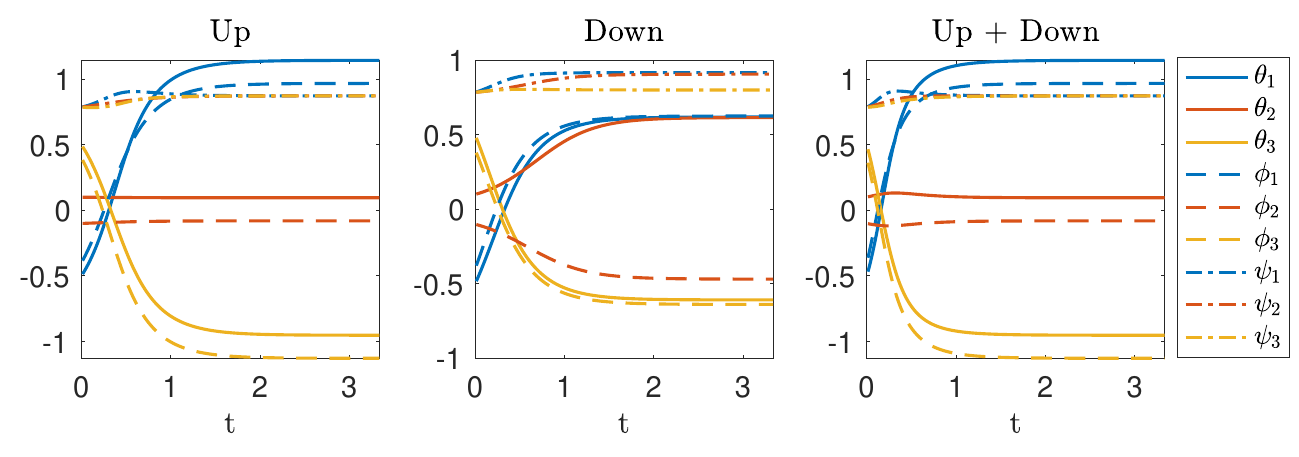}
    \caption{$\delta = \pi/3$}
    \label{fig:diffusion1a}
    \end{subfigure}
    \begin{subfigure}[b]{\textwidth}
    \centering
    \includegraphics[width=\textwidth]{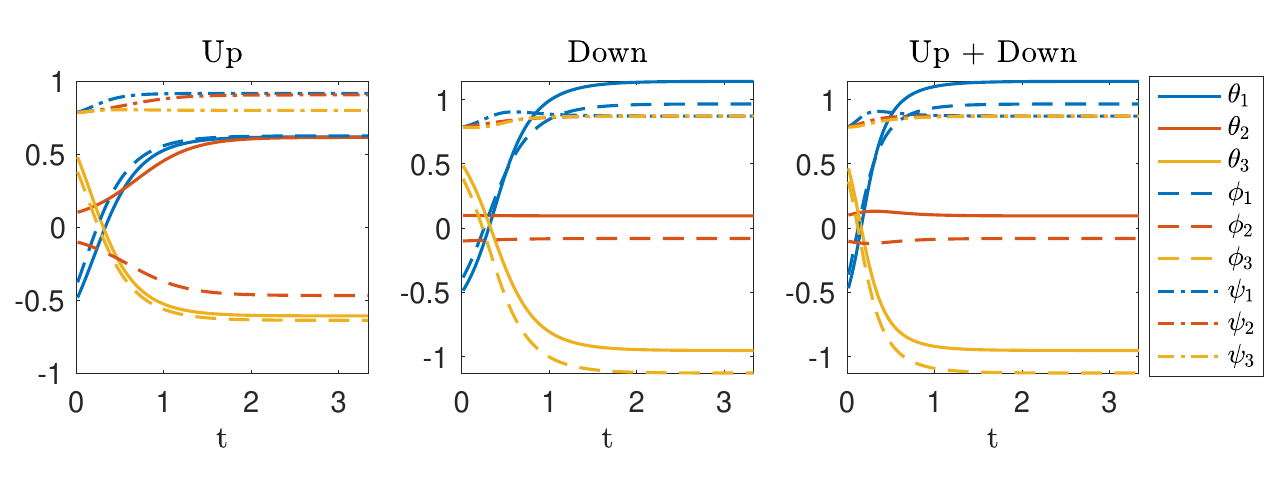}
    \caption{$\delta = 2\pi/3$}
    \label{fig:diffusion1b}
    \end{subfigure}
    \caption{Higher-order diffusion driven by the $1$-Connection Laplacians $\hL_1^{up}$ (Up),  $\hL_1^{down}$ (Down) and $\hL_1$ (Up+Down) on directed simplicial triangle in Case 1 are plotted for single initial conditions and for $\delta=\pi/3$ (upper row) and $\delta=2\pi/3$ (lower row). In the upper-left plot, the final state is an eigenvector of $\hL_{1}^{c, up}$ corresponding to the 0 eigenvalue, while in the lower-middle plot, the equilibrium vector is an eigenvector of $\hL_{1}^{c, down}$ corresponding to the 0 eigenvalue. In the remaining plots, the final states converge to the slowest eigenmodes associated with the smallest positive eigenvalue.}
    \label{fig:diffusion1}
\end{figure}
The $1$-up and the $1$-down Connection Laplacians $\hL_{1}^{c, up}$ and  $\hL_{1}^{c, down}$ commute since the commutator is zero (bottom-right panel in Figure \ref{fig:case1}). Furthermore, it can be shown that $\hL_{1}^{c, down} \hL_{1}^{c, up} \neq 0$, which implies that the Hodge decomposition does not hold, i.e., $\hL_{1}^{c, up}\hL_{1}^{c, down}\neq 0, \hL_{1}^{c, down}\hL_{1}^{c, up}\neq 0$ for all values of $\delta$. 

Figure \ref{fig:diffusion1} plots two examples when the dynamics are described by $\hL_{1}^{c, up}$ (left), $\hL_{1}^{c, down}$ (middle), and their sum (right) for various value of $\delta$. For visualization purposes, we only display the phase angles. Setting $\delta = \pi/3$ (Figure \ref{fig:diffusion1a}) and examining the plot for $\hL_{1}^{c, up}$, we notice that it converges towards an eigenvector of $\hL_{1}^{c, up}$ associated with an eigenvalue of zero, which satisfies conditions (\ref{eq:case1_Lu=0_row1})--(\ref{eq:case1_Lu=0_row3}). Since $\hL_{1}^{c, \text{down}}$ and $\hL_{1}^{c, \text{up}} + \hL_{1}^{c, \text{down}}$ only possess positive eigenvalues when $\delta = \pi/3$, the vectors converge into their slow eigenmodes associated with the smallest eigenvalue. We observe similar results when $\delta = 2\pi/3$ (Figure \ref{fig:diffusion1b}), where the equilibrium vector is an eigenvector of $\hL_{1}^{c, down}$ corresponding to the 0 eigenvalue satisfying 
(\ref{eq:case1_Ld=0_row1})--(\ref{eq:case1_Ld=0_row3}).

\subsubsection{Case 2}
Now, let us consider Case 2 in Table \ref{fig:triangle}, where the edge directions are as follows: $1\to2$, $2\to 3$, and $3\to 1$; and the triangle direction is $3\rightarrow 2\rightarrow 2\rightarrow 1$. This scenario represents a situation where all edge directions are opposite to the direction of the triangle. Consequently, when traversing from one edge to another, we either move against the edge direction while aligning with the triangle direction - as shown in the third row of (\ref{eq:T1u}) - or align with the edge direction while moving against the triangle direction, as depicted in the fourth row of (\ref{eq:T1u}). The corresponding 1-up Connection Laplacian is
\begin{equation}
\hL_{1}^{c, up}=\bordermatrix{
&[1,2]&[1,3]&[2,3]\cr
[1,2]&2{\bm\sigma_0} &-{\bm \sigma}_xe^{-\iu \delta}&{\bm \sigma}_xe^{\iu\delta}\cr
[1,3]&-{\bm \sigma}_xe^{\iu\delta}&2{\bm\sigma_0} &-{\bm \sigma}_xe^{-\iu\delta}\cr
[2,3]&{\bm \sigma}_xe^{-\iu\delta}&-{\bm \sigma}_xe^{\iu\delta}&2{\bm\sigma_0}\cr
}.
\end{equation}
Then the quadratic form $\bnu^H \hL_{1}^{c, up} \bnu$ can be calculated as
\begin{align}
\bnu^H \hL_{1}^{c, up} \bnu &=  \| \nu_1- {\bm \sigma}_xe^{-\iu \delta} \nu_2 \|^2 + \| \nu_1 + {\bm \sigma}_xe^{\iu\delta} \nu_3 \|^2 + \| \nu_2- {\bm \sigma}_xe^{-\iu \delta} \nu_3 \|^2.
\end{align}
We plot its eigenvalues against $\delta$ in the top left panel of Figure (\ref{fig:case1}). The corresponding eigenvalues are:
\begin{eqnarray}
&\left\{ 2 + 2\cos(\delta ), 2 + 2\cos\Big(\delta - \frac{\pi}{3}\Big), 2 + 2\cos\Big(\delta- \frac{2\pi}{3}\Big), 2 + 2\cos(\delta - \pi),  \right.\nonumber \\
&\left. 2 + 2\cos\Big(\delta - \frac{4\pi}{3}\Big), 2 + 2\cos\Big(\delta - \frac{5\pi}{3}\Big) \right\}.
\end{eqnarray}
It may be shown that $\bnu^H \hL_{1}^{c, up} \bnu = 0$ when $\psi_1 = \psi_2 = \psi_3 = \pi/4$, and
\begin{align}
    \label{eq:case2_Lu=0_row1}
    &\theta_1 \equiv \phi_2 - \delta \text{, } \phi_1 \equiv \theta_2 - \delta, \\ 
    &\theta_1 \equiv \phi_3 + \delta + \pi \text{, }  \phi_1 \equiv \theta_3 + \delta + \pi,\\
    &\theta_2 \equiv \phi_3 - \delta \text{, }  \phi_2 \equiv \theta_3 - \delta. 
    \label{eq:case2_Lu=0_row3}
\end{align}
The equations above have a solution only when $\delta \in \{0, \pi/3, 2\pi/3, \pi, 4\pi/3, 5\pi/3 \}$, as can be seen in the plot. In Case 2, $\hL_{1}^{c, down}$ remains the same as in Case 1 since the edge directions stay unchanged. The eigenvalues of $\hL_{1}^{c, up} + \hL_{1}^{c, down}$ are
\begin{equation}
\left\{ 4 + 4\cos\Big(\delta - \frac{\pi}{3}\Big), 4 + 4\cos\Big(\delta - \pi\Big),  4 + 4\cos\big(\delta - \frac{5\pi}{3}\Big), 4 \right\}.
\end{equation}

\begin{figure}
    \centering
    \includegraphics[width=0.9\textwidth]{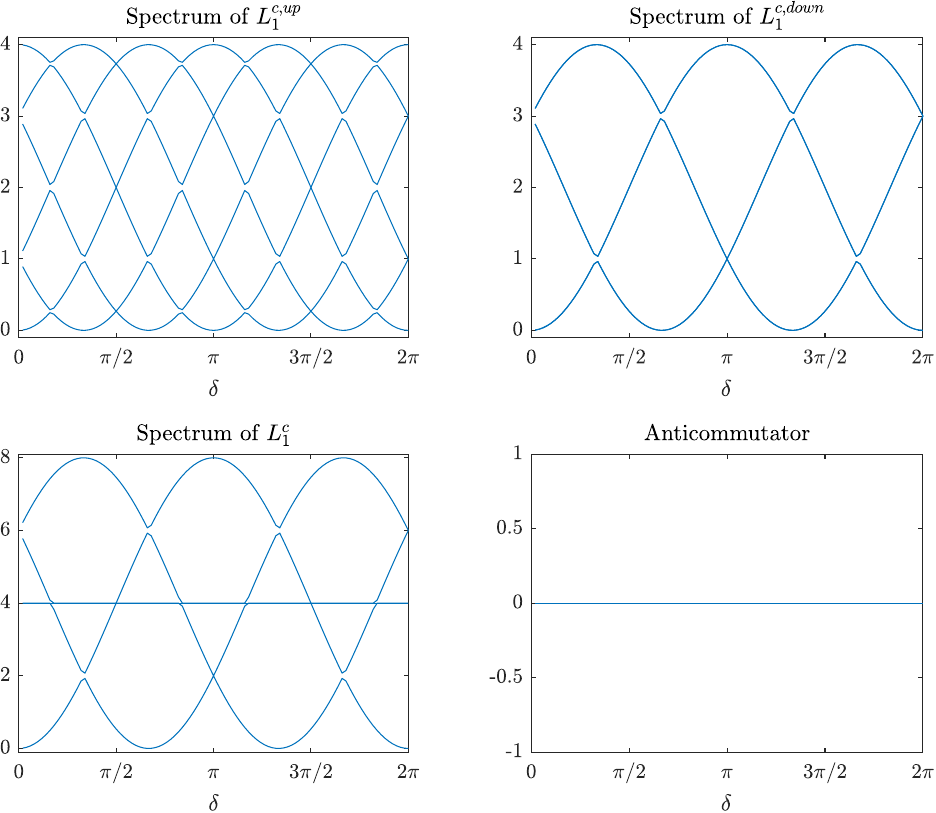}
    \caption{The complete spectra of $\hL_{1}^{c, up}$ (top-left), $\hL_{1}^{c, down}$ (top-right), and $\hL_{1}^{c}$ (bottom-left), and commutator $[\hL_{1}^{c, up}, \hL_{1}^{c, down}]$ (bottom-right)  for Case 2 directed simplicial triangles is plotted as a function of $\delta$.}
    \label{fig:case2}
\end{figure}

\begin{figure}
    \centering
    \begin{subfigure}[b]{\textwidth}
    \centering
    \includegraphics[width=0.85\textwidth]{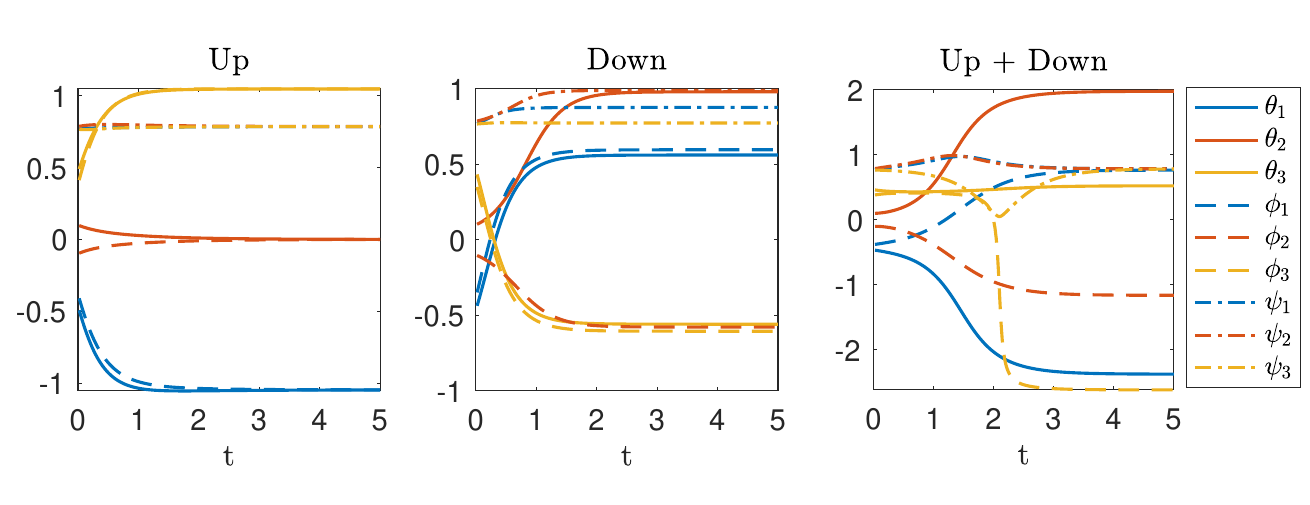}
    \caption{$\delta = \pi/3$}
    \end{subfigure}
    \begin{subfigure}[b]{\textwidth}
    \centering
    \includegraphics[width=0.85\textwidth]{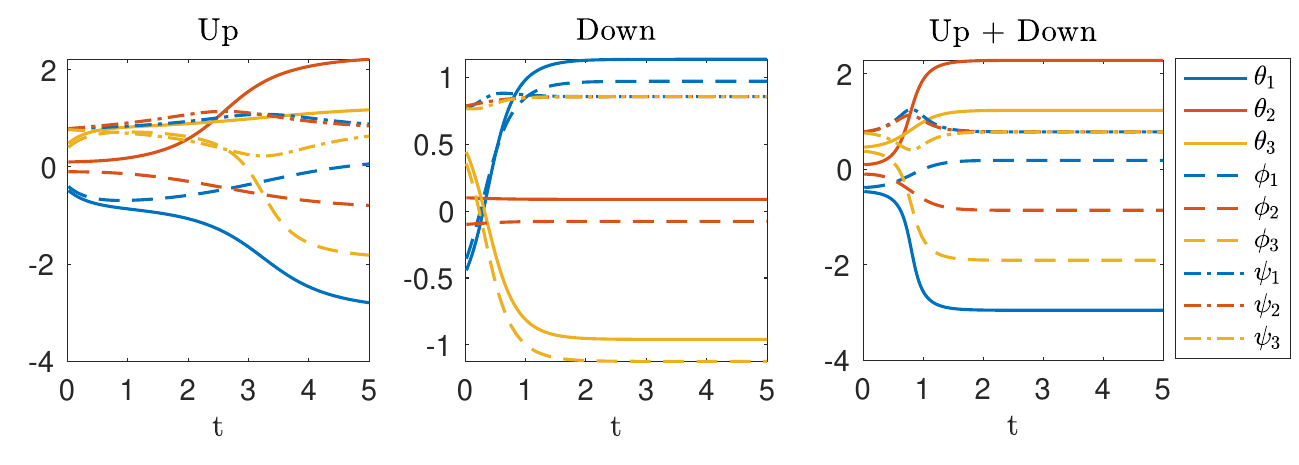}
    \caption{$\delta = 2\pi/3$}
    \end{subfigure}
    \caption{Higher-order diffusion driven by the $1$-Connection Laplacians $\hL_1^{up}$ (Up), $\hL_1^{down}$ (Down) and $\hL_1$ (Up+Down) on directed simplicial triangle in Case 2 are plotted for single initial conditions and for $\delta=\pi/3$ (upper row) and $\delta=2\pi/3$ (lower row). In the upper-left plot, vectors converge towards an eigenvector of $\hL_{1}^{c, \text{up}}$ with an eigenvalue of zero. In the lower plot, where $\delta = 2\pi/3$, both equilibrium vectors for $\hL_{1}^{c, \text{up}}$ (on the left) and $\hL_{1}^{c, \text{down}}$ (in the middle) are eigenvectors associated with a zero eigenvalue. Across the remaining plots, the final states converge to the slowest eigenmodes linked to the smallest positive eigenvalue due to the absence of a zero eigenvalue.}
    \label{fig:diffusion2}
\end{figure}

The commutator $[\hL_{1}^{c, down},\hL_{1}^{c, up}]$ is zero, i.e., $[\hL_{1}^{c, down},\hL_{1}^{c, up}]=0$, as shown in the bottom-right of Figure~\ref{fig:case2}, however Hodge decomposition does not hold as in Case 1. Figure \ref{fig:diffusion2} displays two examples of vector diffusion. When $\delta = \pi/3$, the vectors in the left plot converge toward an eigenvector of $\hL_{1}^{c, \text{up}}$ corresponding to eigenvalue zero, satisfying conditions (\ref{eq:case2_Lu=0_row1})--(\ref{eq:case2_Lu=0_row3}). When $\delta = 2\pi/3$, the equilibrium vector for both $\hL_{1}^{c, \text{up}}$ (on the left) and $\hL_{1}^{c, \text{down}}$ (in the middle) are eigenvectors associated with a zero eigenvalue, as specified by conditions (\ref{eq:case2_Lu=0_row1})--(\ref{eq:case2_Lu=0_row3}) and 
(\ref{eq:case1_Ld=0_row1})--(\ref{eq:case1_Ld=0_row3}) respectively. In the remaining plots, the final states converge toward the slowest eigenmodes associated with the smallest positive eigenvalue due to the absence of a zero eigenvalue.

\subsubsection{Case 3}
\begin{figure}
\centering
\includegraphics[width=0.9\textwidth]{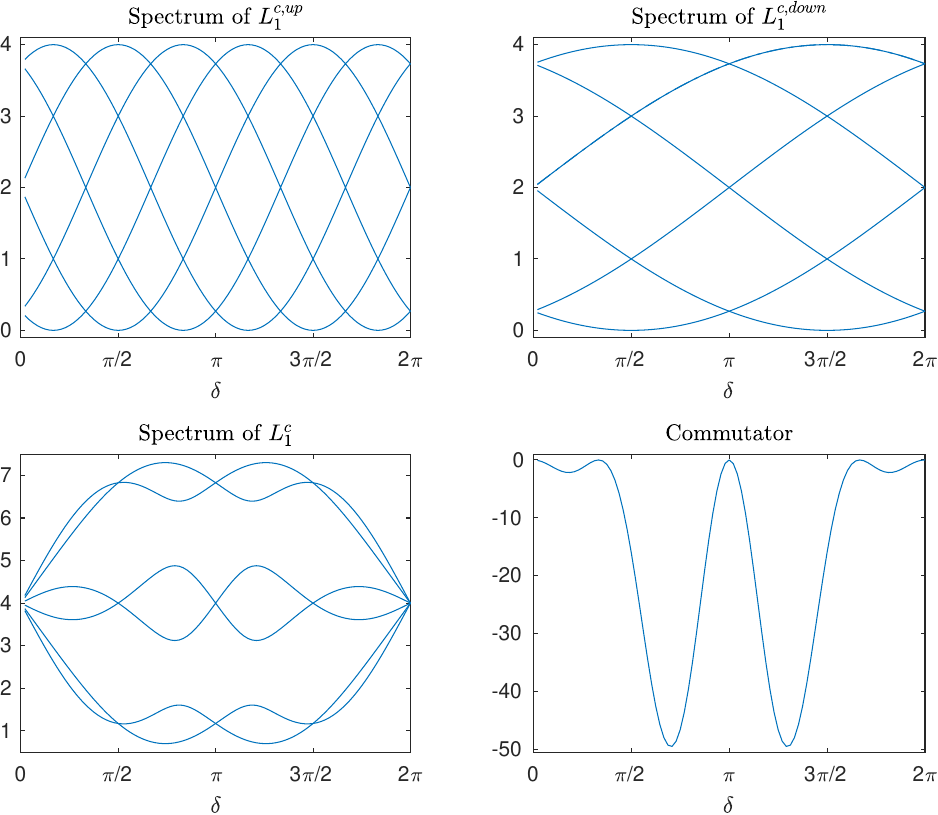}
\caption{The complete spectra of $1$-Connection Laplacians $\hL_{1}^{c, up}$ (top-left), $\hL_{1}^{c, down}$ (top-right), and $\hL_{1}^{c}$ (bottom-left), and commutator $[\hL_{1}^{c, up}, \hL_{1}^{c, down}]$(bottom-right) for Case 3 directed simplicial triangles is plotted as a function of $\delta$.}
\label{fig:case3}
\end{figure}

Now, let us look at the case of a $2$-simplicial complex with triangle and link direction given by $1\rightarrow 2\rightarrow 3$ and $1\to2$, $2\to 3$ and $1\to 3$, which is illustrated as Case 3 in Table \ref{fig:triangle}. First, we can write the 1-up and 1-down Connection  Laplacian as
\begin{equation}
\hL_{1}^{c, up}=\bordermatrix{
&[1,2]&[1,3]&[2,3]\cr
[1,2]& 2{\bm\sigma_0}  & -{\bm \sigma}_y e^{\iu\delta}&  {\bm\sigma_0}  e^{-\iu\delta}\cr
[1,3]&-{\bm \sigma}_y e^{-\iu\delta}& 2{\bm\sigma_0} &-{\bm \sigma}_ze^{\iu\delta}\cr
[2,3]& {\bm\sigma_0}  e^{\iu\delta}&-{\bm \sigma}_ze^{-\iu\delta}& 2{\bm\sigma_0} \cr
}.
\end{equation}
Its quadratic form is 
\begin{align}
\bnu^H \hL_{1}^{c, up} \bnu &=  \| \nu_1- {\bm \sigma}_ye^{\iu \delta} \nu_2 \|^2 + \| \nu_1 + {\bm\sigma_0}e^{-\iu\delta} \nu_3 \|^2 + \| \nu_2- {\bm \sigma}_z e^{\iu \delta} \nu_3 \|^2,
\end{align}
and it becomes zero when $\psi_1 = \psi_2 = \psi_3 = \pi/4$ and
\begin{align}
    \label{eq:case3_Lu=0_row1} 
    &\theta_1 \equiv \phi_2 + \delta - \frac{\pi}{2} \text{, } \phi_1 \equiv \theta_2 + \delta +\frac{\pi}{2}, \\  
    &\theta_1 \equiv \theta_3 - \delta + \pi \text{, }  \phi_1 \equiv \phi_3 - \delta + \pi,\\    
    &\theta_2 \equiv \phi_3 + \delta \text{, }  \phi_2 \equiv \phi_3 + \delta + \pi. \label{eq:case3_Lu=0_row3}
\end{align}
The above system has a solution iff $\delta \in \{\pi/6, \pi/2, 5\pi/6, 7\pi/6, 3\pi/2, 11\pi/6 \}$. This can be confirmed by the eigenvalue top-left plot in Figure \ref{fig:case3}. Furthermore, the eigenvalues are given by 
\begin{eqnarray}
&\left\{ 2 + 2\cos\Big(\delta - \frac{\pi}{6}\Big), 2 + 2\cos\Big(\delta - \frac{\pi}{2}\Big),\right. \nonumber \\ &2 + 2\cos\Big(\delta- \frac{5\pi}{6}\Big), 2 + 2\cos\Big(\delta - \frac{7\pi}{6}\Big),  \nonumber \\
&\left. 2 + 2\cos\Big(\delta - \frac{3\pi}{2}\Big), 2 + 2\cos\Big(\delta - \frac{11\pi}{6}\Big) \right\}.
\end{eqnarray}
For the $1$-down Connection Laplacian, we have
\begin{equation}
\hL_{1}^{c, down}=\bordermatrix{
&[1,2]&[1,3]&[2,3]\cr
[1,2]&2{\sigma_0} & {\bm \sigma}_y &-{\bm\sigma_0} e^{\iu\delta}\cr
[1,3]&{\bm \sigma}_y&2{\bm\sigma_0} &{\bm \sigma}_z\cr
[2,3]&-{\bm\sigma_0} e^{-\iu\delta}&{\bm \sigma}_z&2{\bm\sigma_0}\cr },
\end{equation}
and
\begin{align}
\bnu^H \hL_{1}^{c, down} \bnu &=  \| \nu_1+ {\bm \sigma}_y\nu_2 \|^2 + \| \nu_1 - {\bm\sigma_0}e^{\iu\delta} \nu_3 \|^2 + \| \nu_2+ {\bm \sigma}_z  \nu_3 \|^2.
\end{align}
The quadratic form is zero when $\psi_1 = \psi_2 = \psi_3 = \pi/4$ and 
\begin{align}
   \label{eq:case3_Ld=0_row1} &\theta_1 \equiv \phi_2 + \frac{\pi}{2} \text{, } \phi_1 \equiv \theta_2 - \frac{\pi}{2}, \\ 
    &\theta_1 \equiv \theta_3 + \delta  \text{, }  \phi_1 \equiv \phi_3 + \delta ,\\    
    &\theta_2 \equiv \theta_3 + \pi \text{, }  \phi_2 \equiv \phi_3.     \label{eq:case3_Ld=0_row3}
\end{align}
The equation above can be solved iff $\delta \in \{\pi/2, 3\pi/2\}$ as shown in the top-right plot in Figure \ref{fig:case3}. Furthermore, the eigenvalues are given by 
\begin{eqnarray}
& \left\{ 2 + 2\cos\Big(\frac{\delta}{3} - \frac{\pi}{6}\Big), 2 - 2\cos\Big(\frac{\delta}{3} - \frac{\pi}{6}\Big), \right.\nonumber\\& 2 + 2\cos\Big(\frac{\delta}{3}- \frac{\pi}{2}\Big), 2 - 2\cos\Big(\frac{\delta}{3} - \frac{\pi}{2}\Big),  \nonumber\\
& \left. 2 + 2\cos\Big(\frac{\delta}{3} - \frac{5\pi}{6}\Big), 2 - 2\cos\Big(\frac{\delta}{3} - \frac{5\pi}{6}\Big) \right\}.
\end{eqnarray}

\begin{figure}
\centering
\begin{subfigure}[b]{\textwidth}
\centering
\includegraphics[width=0.85\textwidth]{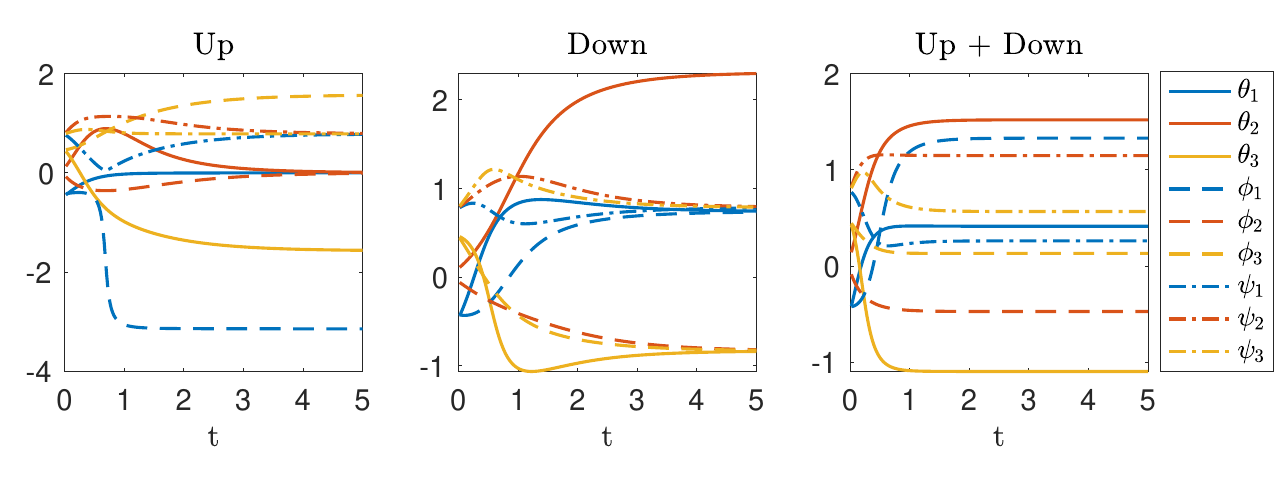}
\caption{$\delta = \pi/2$}
\end{subfigure}
\begin{subfigure}[b]{\textwidth}
\centering
\includegraphics[width=0.85\textwidth]{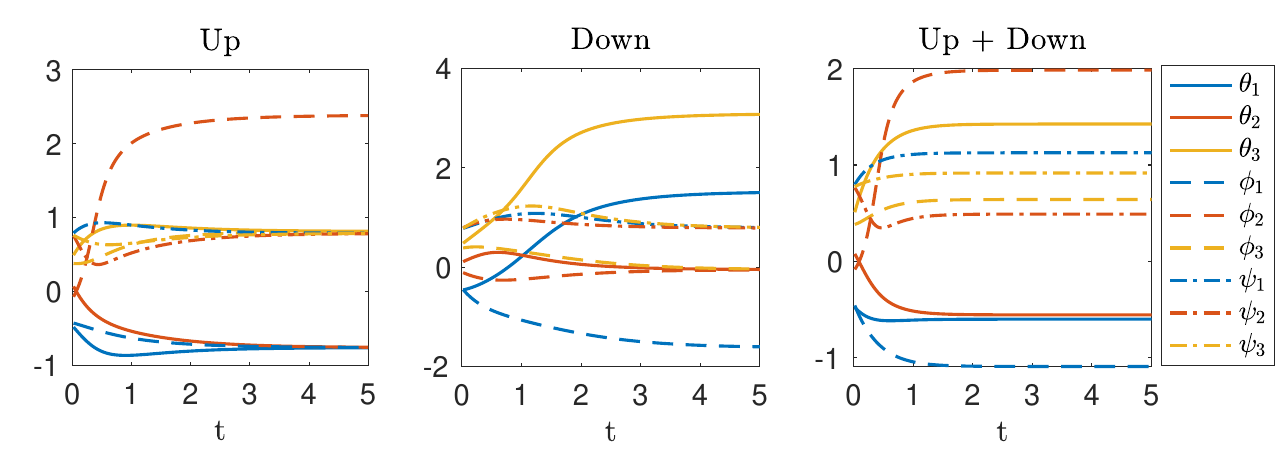}
\caption{$\delta = 3\pi/2$}
\end{subfigure}
\caption{Higher-order diffusion driven by the $1$-Connection Laplacians $\hL_1^{up}$ (Up), $\hL_1^{down}$ (Down) and $\hL_1$ (Up+Down) on directed simplicial triangle in Case 3 are plotted for single initial conditions and for $\delta=\pi/3$ (upper row) and $\delta=2\pi/3$ (lower row). In both situations, the vectors associated with $\hL_{1}^{c, \text{up}}$ and $\hL_{1}^{c, \text{down}}$ converge towards eigenvectors corresponding to a zero eigenvalue.}
\label{fig:diffusion3}
\end{figure}

In this case $\hL_{1}^{c, up} $ and $\hL_{1}^{c, up}$ do not commute except for the special case when $\delta = 0$ as shown in the bottom-right panel in Figure \ref{fig:case3}. Two examples of vector diffusion are displayed in Figure \ref{fig:diffusion3} for $\delta = \pi/2$ and $3\pi/2$. In both scenarios, the vectors for $\hL_{1}^{c, up} $ and $\hL_{1}^{c, down}$, the vectors approach some eigenvectors that correspond to eigenvalue zero, which satisfies conditions (\ref{eq:case3_Lu=0_row1})-(\ref{eq:case3_Lu=0_row3}) and (\ref{eq:case3_Ld=0_row1})-(\ref{eq:case3_Ld=0_row3}) respectively.

\subsubsection{Case 4}
\begin{figure}
\centering
\includegraphics[width=0.9\textwidth]{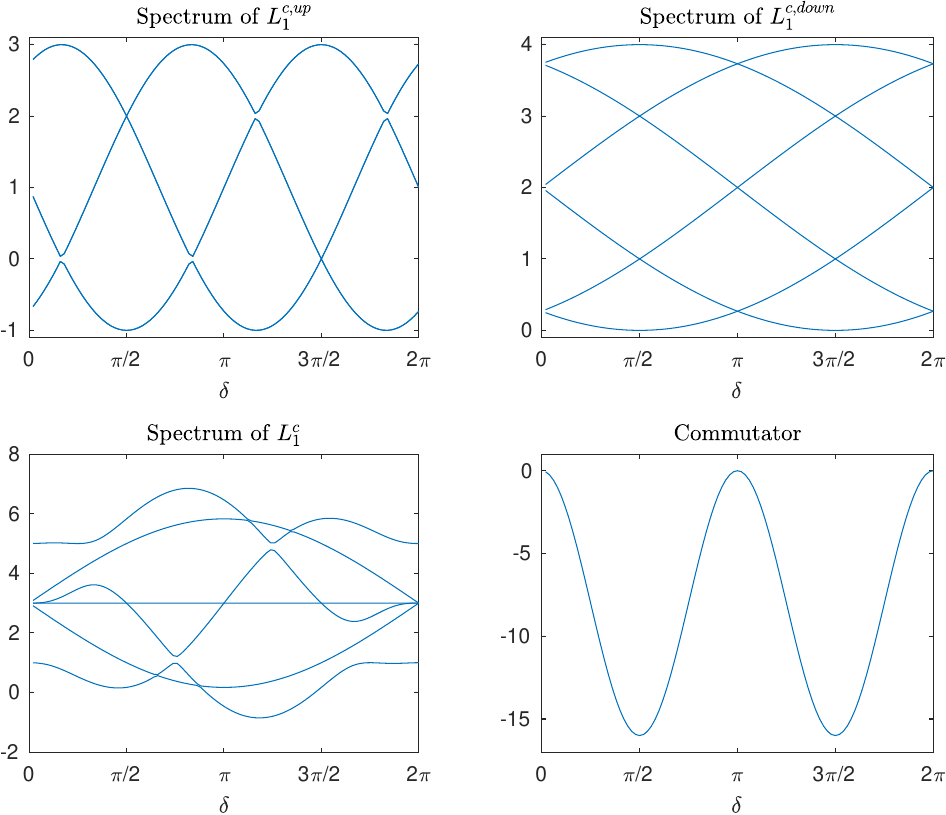}
\caption{The complete spectra of the $1$-Connection Laplacians $\hL_{1}^{c, up}$ (top-left), $\hL_{1}^{c, down}$ (top-right), and $\hL_{1}^{c}$ (bottom-left), and commutator $[\hL_{1}^{c, up}, \hL_{1}^{c, down}] $(bottom-right) for Case 4 of directed simplicial triangles is plotted as a function of $\delta$.}
\label{fig:case4}
\end{figure}

\begin{figure}
\centering
\begin{subfigure}[b]{\textwidth}
\centering
\includegraphics[width=\textwidth]{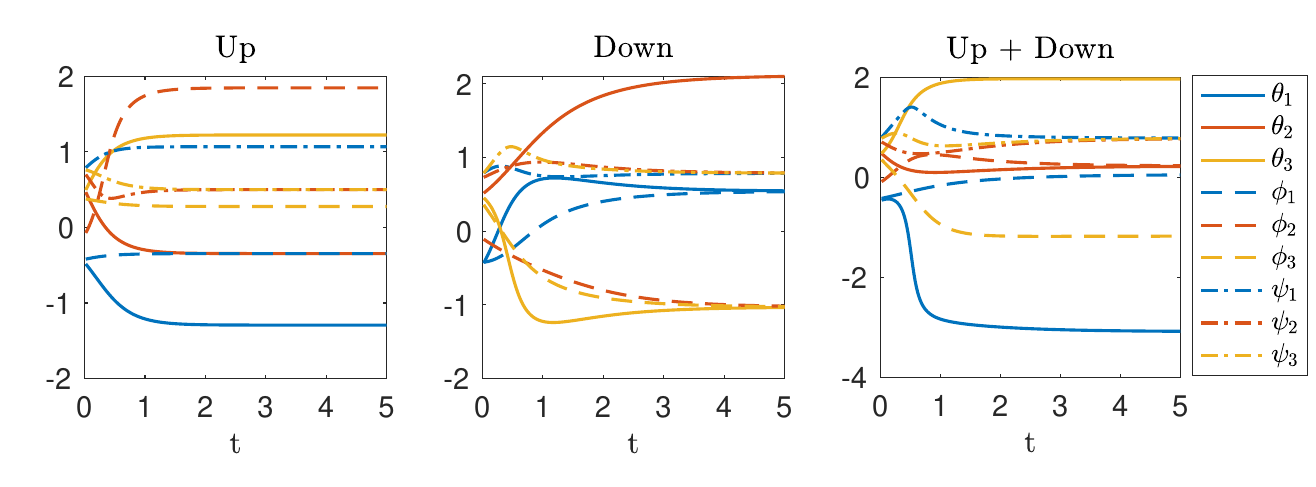}
\caption{$\delta = \pi/2$}
\end{subfigure}
\begin{subfigure}[b]{\textwidth}
\centering
\includegraphics[width=\textwidth]{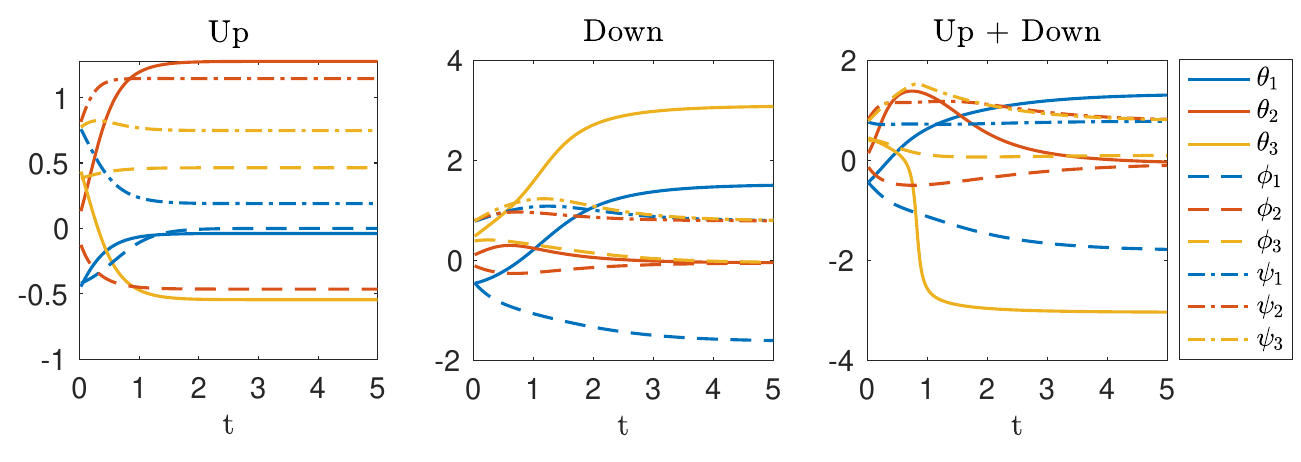}
\caption{$\delta = 3\pi/2$}
\end{subfigure}
\caption{Higher-order diffusion driven by the $1$-Connection Laplacians $\hL_1^{up}$ (Up), $\hL_1^{down}$ (Down) and $\hL_1$ (Up+Down) on directed simplicial triangle in Case 4 are plotted for single initial conditions. When $\delta=\pi/3$ (top row), the equilibrium vector coincides with the eigenvector linked to a zero eigenvalue for both $\hL_{1}^{c, up}$ and $\hL_{1}^{c, down}$. Conversely, when $\delta=2\pi/3$ (bottom row), only the vector associated with $\hL_{1}^{c, down}$ converges towards the eigenvector with a zero eigenvalue. In the remaining scenarios, the final state is determined by the vector possessing the smallest positive eigenvalue.}
\label{fig:diffusion4}
\end{figure}

Finally, we examine the directed $2$-simplicial complex with triangle and link direction given by $3\rightarrow 2\rightarrow 1$ and $1\to2$, $2\to 3$ and $1\to 3$ illustrated as Case 4 in Figure \ref{fig:triangle}. As the edge directions are precisely the same as in Case 3, $\hL_{1}^{c, down}$ in this scenario is identical to the previous case. Therefore, we only need to discuss $\hL_{1}^{c, up}$, which can be computed as follows:
\begin{equation}
\hL_{1}^{c, up}=\bordermatrix{
&[1,2]&[1,3]&[2,3]\cr
[1,2]&{2\bm\sigma_0} &-{\bm \sigma}_y e^{-\iu\delta}&{\bm \sigma}_xe^{\iu\delta}\cr
[1,3]&-{\bm \sigma}_y e^{\iu\delta}&{2\bm\sigma_0} &-{\bm \sigma}_ze^{-\iu\delta}\cr
[2,3]&{\bm \sigma}_xe^{-\iu\delta}&-{\bm \sigma}_ze^{\iu\delta}&{2\bm\sigma_0}\cr }.
\end{equation}
This leads to 
\begin{align}
\bnu^H \hL_{1}^{c, up} \bnu &=  \| \nu_1 - {\bm \sigma}_y e^{-\iu\delta} \nu_2 \|^2 + \| \nu_1 + {\bm \sigma}_x e^{\iu\delta} \nu_3 \|^2 + \| \nu_2 - {\bm \sigma}_z e^{-\iu\delta} \nu_3 \|^2.
\end{align}
We can further prove that $\bnu^H \hL_{1}^{c, up} \bnu = 0$ when $\pi/2 - \psi_1 = \psi_2 = \psi_3$, and
\begin{align}
    &\theta_1 \equiv \phi_2 - \delta - \frac{\pi}{2} \text{, } \phi_1 \equiv \theta_2 - \delta +\frac{\pi}{2}, \\
    &\theta_1 \equiv \phi_3 + \delta + \pi \text{, }  \phi_1 \equiv \theta_3 + \delta + \pi,\\    
    &\theta_2 \equiv \phi_3 - \delta \text{, }  \phi_2 \equiv \phi_3 - \delta + \pi.   
\end{align}
The solution to the equations exists iff $\delta \in \{ \pi/2, 7\pi/6, 11\pi/6\}$ as shown in the top-left plot in Figure \ref{fig:case4}. The eigenvalues shown in the top-left subplot are:
\begin{equation}
\left\{ 2 + 2\cos\Big(\delta + \frac{\pi}{2}\Big), 2 + 2\cos\Big(\delta - \frac{5\pi}{6}\Big),2 + 2\cos\Big(\delta - \frac{\pi}{6}\Big) \right\}.
\end{equation}Furthermore, we observe that $\hL_{1}^{c, up}$ and $\hL_{1}^{c, down}$ only commute when $\delta = 0$ or $\pi$. In Figure \ref{fig:diffusion4}, we illustrate two vector diffusion processes where $\delta = \pi$ and $3\pi/2$. In the top row, the equilibrium vector is the eigenvector associated with a zero eigenvalue for both $\hL_{1}^{c, up}$ and $\hL_{1}^{c, down}$. In the bottom plot, only the vector for $\hL_{1}^{c, down}$ converges to the eigenvector with a zero eigenvalue. For the remaining cases, the final state corresponds to the vector with the smallest positive eigenvalue, given the absence of a zero eigenvalue. 

\subsection{Case Study on Triangulated Torus}
\label{chp3:torus}
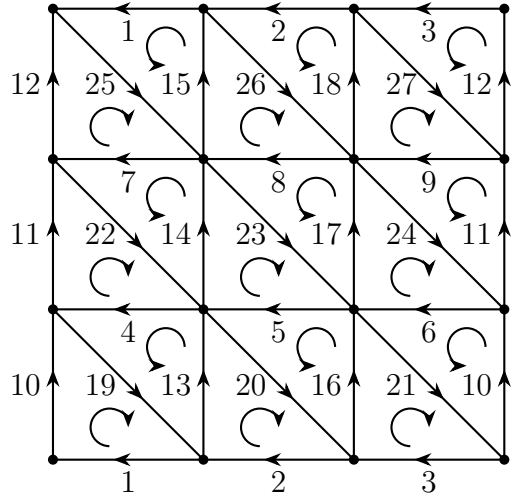
\begin{figure}[ht]
\centering
\begin{tikzpicture}[>=Stealth, thick, decoration={
    markings,
    mark=at position 0.6 with \arrow{>}}
    ] 
\draw[fill=black] (0,0) circle (1.5pt);
\draw[fill=black] (0,2) circle (1.5pt);
\draw[fill=black] (0,4) circle (1.5pt);
\draw[fill=black] (0,6) circle (1.5pt);
\draw[fill=black] (2,0) circle (1.5pt);
\draw[fill=black] (2,2) circle (1.5pt);
\draw[fill=black] (2,4) circle (1.5pt);
\draw[fill=black] (2,6) circle (1.5pt);
\draw[fill=black] (4,0) circle (1.5pt);
\draw[fill=black] (4,2) circle (1.5pt);
\draw[fill=black] (4,4) circle (1.5pt);
\draw[fill=black] (4,6) circle (1.5pt);
\draw[fill=black] (6,0) circle (1.5pt);
\draw[fill=black] (6,2) circle (1.5pt);
\draw[fill=black] (6,4) circle (1.5pt);
\draw[fill=black] (6,6) circle (1.5pt);

\draw[postaction={decorate}](2, 0) -- node[below=0.1mm] {1} (0, 0); \draw[postaction={decorate}] (4, 0) -- node[below=0.1mm] {2}(2, 0); \draw[postaction={decorate}] (6, 0) -- node[below=0.1mm] {3}(4, 0); 
\draw[postaction={decorate}] (2, 2) -- node[below=0.1mm] {4}(0, 2); \draw[postaction={decorate}] (4, 2) -- node[below=0.1mm] {5}(2, 2); \draw[postaction={decorate}] (6, 2) -- node[below=0.1mm] {6}(4, 2); 
\draw[postaction={decorate}] (2, 4) -- node[below=0.1mm] {7} (0, 4); \draw[postaction={decorate}] (4, 4) -- node[below=0.1mm] {8} (2, 4); \draw[postaction={decorate}] (6, 4) -- node[below=0.1mm] {9}(4, 4); 
\draw[postaction={decorate}] (2, 6) -- node[below=0.1mm] {1} (0, 6); \draw[postaction={decorate}] (4, 6) -- node[below=0.1mm] {2} (2, 6); \draw[postaction={decorate}] (6, 6) -- node[below=0.1mm] {3} (4, 6); 
\draw[postaction={decorate}] (0, 0) -- node[left=0.1mm] {10} (0, 2); \draw[postaction={decorate}] (0, 2) -- node[left=0.1mm] {11} (0, 4); \draw[postaction={decorate}] (0, 4) -- node[left=0.1mm] {12} (0, 6);
\draw[postaction={decorate}] (2, 0) -- node[left=0.1mm] {13} (2, 2); \draw[postaction={decorate}] (2, 2) -- node[left=0.1mm] {14} (2, 4); \draw[postaction={decorate}] (2, 4) -- node[left=0.1mm] {15} (2, 6);
\draw[postaction={decorate}] (4, 0) -- node[left=0.1mm] {16} (4, 2); \draw[postaction={decorate}] (4, 2) -- node[left=0.1mm] {17} (4, 4); \draw[postaction={decorate}] (4, 4) -- node[left=0.1mm] {18} (4, 6);
\draw[postaction={decorate}] (6, 0) -- node[left=0.1mm] {10} (6, 2); \draw[postaction={decorate}] (6, 2) -- node[left=0.1mm] {11} (6, 4); \draw[postaction={decorate}] (6, 4) -- node[left=0.1mm] {12} (6, 6);
\draw[postaction={decorate}] (0, 2) -- node[left=0.1mm] {19} (2, 0); \draw[postaction={decorate}] (2, 2) -- node[left=0.1mm] {20} (4, 0); \draw[postaction={decorate}] (4, 2) -- node[left=0.1mm] {21} (6, 0);
\draw[postaction={decorate}] (0, 4) -- node[left=0.1mm] {22} (2, 2); \draw[postaction={decorate}] (2, 4) -- node[left=0.1mm] {23} (4, 2); \draw[postaction={decorate}] (4, 4) -- node[left=0.1mm] {24} (6, 2);
\draw[postaction={decorate}] (0, 6) -- node[left=0.1mm] {25} (2, 4); \draw[postaction={decorate}] (2, 6) -- node[left=0.1mm] {26} (4, 4); \draw[postaction={decorate}] (4, 6) -- node[left=0.1mm] {27} (6, 4);

\draw[thick, <-] (1,0.43) arc (0:270:0.25); \draw[thick, <-] (3,0.43) arc (0:270:0.25);\draw[thick, <-] (5,0.43) arc (0:270:0.25);
\draw[thick, <-] (1,2.43) arc (0:270:0.25);\draw[thick, <-] (3,2.43) arc (0:270:0.25); \draw[thick, <-] (5,2.43) arc (0:270:0.25);
\draw[thick, <-] (1,4.43) arc (0:270:0.25);\draw[thick, <-] (3,4.43) arc (0:270:0.25); \draw[thick, <-] (5,4.43) arc (0:270:0.25);

\draw[thick, ->] (1.75,1.5) arc (0:270:0.25); \draw[thick, ->] (3.75,1.5) arc (0:270:0.25); \draw[thick, ->] (5.75,1.5) arc (0:270:0.25);
\draw[thick, ->] (1.75,3.5) arc (0:270:0.25); \draw[thick, ->] (3.75,3.5) arc (0:270:0.25); \draw[thick, ->] (5.75,3.5) arc (0:270:0.25);
\draw[thick, ->] (1.75,5.5) arc (0:270:0.25); \draw[thick, ->] (3.75,5.5) arc (0:270:0.25); \draw[thick, ->] (5.75,5.5) arc (0:270:0.25);

\end{tikzpicture}
\caption{Example of a Type 1 Triangulated Torus  of size $3\times 3$ having $25$ edges. In this case, there is no frustration as all edge directions align with the triangle directions.}
\label{fig:torus1}
\end{figure}

\begin{figure}[ht]
\centering
\begin{tikzpicture}[>=Stealth, thick, decoration={
    markings,
    mark=at position 0.6 with \arrow{>}}
    ] 
\draw[fill=black] (0,0) circle (1.5pt);
\draw[fill=black] (0,2) circle (1.5pt);
\draw[fill=black] (0,4) circle (1.5pt);
\draw[fill=black] (0,6) circle (1.5pt);
\draw[fill=black] (2,0) circle (1.5pt);
\draw[fill=black] (2,2) circle (1.5pt);
\draw[fill=black] (2,4) circle (1.5pt);
\draw[fill=black] (2,6) circle (1.5pt);
\draw[fill=black] (4,0) circle (1.5pt);
\draw[fill=black] (4,2) circle (1.5pt);
\draw[fill=black] (4,4) circle (1.5pt);
\draw[fill=black] (4,6) circle (1.5pt);
\draw[fill=black] (6,0) circle (1.5pt);
\draw[fill=black] (6,2) circle (1.5pt);
\draw[fill=black] (6,4) circle (1.5pt);
\draw[fill=black] (6,6) circle (1.5pt);

\draw[postaction={decorate}](2, 0) -- node[below=0.1mm] {1} (0, 0); \draw[postaction={decorate}] (4, 0) -- node[below=0.1mm] {2}(2, 0); \draw[postaction={decorate}] (6, 0) -- node[below=0.1mm] {3}(4, 0); 
\draw[postaction={decorate}] (2, 2) -- node[below=0.1mm] {4}(0, 2); \draw[postaction={decorate}] (4, 2) -- node[below=0.1mm] {5}(2, 2); \draw[postaction={decorate}] (6, 2) -- node[below=0.1mm] {6}(4, 2); 
\draw[postaction={decorate}] (2, 4) -- node[below=0.1mm] {7} (0, 4); \draw[postaction={decorate}] (4, 4) -- node[below=0.1mm] {8} (2, 4); \draw[postaction={decorate}] (6, 4) -- node[below=0.1mm] {9}(4, 4); 
\draw[postaction={decorate}] (2, 6) -- node[below=0.1mm] {1} (0, 6); \draw[postaction={decorate}] (4, 6) -- node[below=0.1mm] {2} (2, 6); \draw[postaction={decorate}] (6, 6) -- node[below=0.1mm] {3} (4, 6); 
\draw[postaction={decorate}] (0, 0) -- node[left=0.1mm] {10} (0, 2); \draw[postaction={decorate}] (0, 2) -- node[left=0.1mm] {11} (0, 4); \draw[postaction={decorate}] (0, 4) -- node[left=0.1mm] {12} (0, 6);
\draw[postaction={decorate}] (2, 0) -- node[left=0.1mm] {13} (2, 2); \draw[postaction={decorate}] (2, 2) -- node[left=0.1mm] {14} (2, 4); \draw[postaction={decorate}] (2, 4) -- node[left=0.1mm] {15} (2, 6);
\draw[postaction={decorate}] (4, 0) -- node[left=0.1mm] {16} (4, 2); \draw[postaction={decorate}] (4, 2) -- node[left=0.1mm] {17} (4, 4); \draw[postaction={decorate}] (4, 4) -- node[left=0.1mm] {18} (4, 6);
\draw[postaction={decorate}] (6, 0) -- node[left=0.1mm] {10} (6, 2); \draw[postaction={decorate}] (6, 2) -- node[left=0.1mm] {11} (6, 4); \draw[postaction={decorate}] (6, 4) -- node[left=0.1mm] {12} (6, 6);
\draw[postaction={decorate}] (0, 2) -- node[left=0.1mm] {19} (2, 0); \draw[postaction={decorate}] (2, 2) -- node[left=0.1mm] {20} (4, 0); \draw[postaction={decorate}] (4, 2) -- node[left=0.1mm] {21} (6, 0);
\draw[postaction={decorate}] (0, 4) -- node[left=0.1mm] {22} (2, 2); \draw[postaction={decorate}] (2, 4) -- node[left=0.1mm] {23} (4, 2); \draw[postaction={decorate}] (4, 4) -- node[left=0.1mm] {24} (6, 2);
\draw[postaction={decorate}] (0, 6) -- node[left=0.1mm] {25} (2, 4); \draw[postaction={decorate}] (2, 6) -- node[left=0.1mm] {26} (4, 4); \draw[postaction={decorate}] (4, 6) -- node[left=0.1mm] {27} (6, 4);

\draw[thick, <-] (1,0.43) arc (0:270:0.25); \draw[thick, <-] (3,0.43) arc (0:270:0.25);\draw[thick, <-] (5,0.43) arc (0:270:0.25);
\draw[thick, <-] (1,2.43) arc (0:270:0.25);\draw[thick, <-] (3,2.43) arc (0:270:0.25); \draw[thick, <-] (5,2.43) arc (0:270:0.25);
\draw[thick, <-] (1,4.43) arc (0:270:0.25);\draw[thick, <-] (3,4.43) arc (0:270:0.25); \draw[thick, <-] (5,4.43) arc (0:270:0.25);

\draw[thick, <-] (1.75,1.5) arc (0:270:0.25); \draw[thick, <-] (3.75,1.5) arc (0:270:0.25); \draw[thick, <-] (5.75,1.5) arc (0:270:0.25);
\draw[thick, <-] (1.75,3.5) arc (0:270:0.25); \draw[thick, <-] (3.75,3.5) arc (0:270:0.25); \draw[thick, <-] (5.75,3.5) arc (0:270:0.25);
\draw[thick, <-] (1.75,5.5) arc (0:270:0.25); \draw[thick, <-] (3.75,5.5) arc (0:270:0.25); \draw[thick, <-] (5.75,5.5) arc (0:270:0.25);

\end{tikzpicture}
\caption{Example of a Type 2 Triangulated Torus  of size $3\times 3$ having $25$ edges. The flow direction is reversed for all upper triangles above the diagonal edges, as compared to the  Type 1 Triangulated Torus.} 
\label{fig:torus2}
\end{figure}
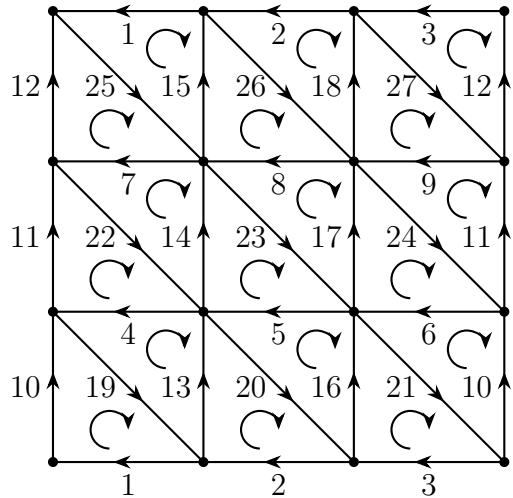

In the previous examples, we only considered simplicial complexes with a single triangle. Now, we will examine examples where the simplicial complexes contain multiple triangles and edges. Specifically, we will focus on the triangulated torus as our example. The torus is a 2-manifold without boundary. To perform computations, we triangulate the torus into a simplicial complex and we consider two cases that differ only in the direction of the triangles (see Figure $\ref{fig:torus1}$,  and Figure $\ref{fig:torus2}$). For each case, we will investigate the spectrum and the complete spectrum of the 1-up and 1-down Connection Laplacians and we provide a discussion of the effects of the frustration induced by the directions of the simplices.

We first consider the Type 1  Triangulated Torus  illustrated in Figure \ref{fig:torus1}. In this case, all edge directions align with the triangle directions. Therefore the triangles fall under Case 1 of directed triangles in Figure \ref{fig:triangle}. 



We then reverse the direction of the upper triangles above the diagonal edges to define the Type 2 Triangulated Torus illustrated in Figure $\ref{fig:torus2}$. As a result, these upper triangles align with Case 2 in Figure \ref{fig:triangle} while the lower triangles correspond to Case 1. Note that for both the Type 1 and the Type 2 Torus we adopt the usual convention for defining the orientations of the edges of their undirected counterpart: instead of taking the orientations induced by the node labels we consider orientations that are consistent with periodic boundary conditions that in these cases are aligned to the direction of the edges depicted in Figures $\ref{fig:torus1}$ and $\ref{fig:torus2}$.

\begin{figure}
\centering
\includegraphics[width=0.9\textwidth]{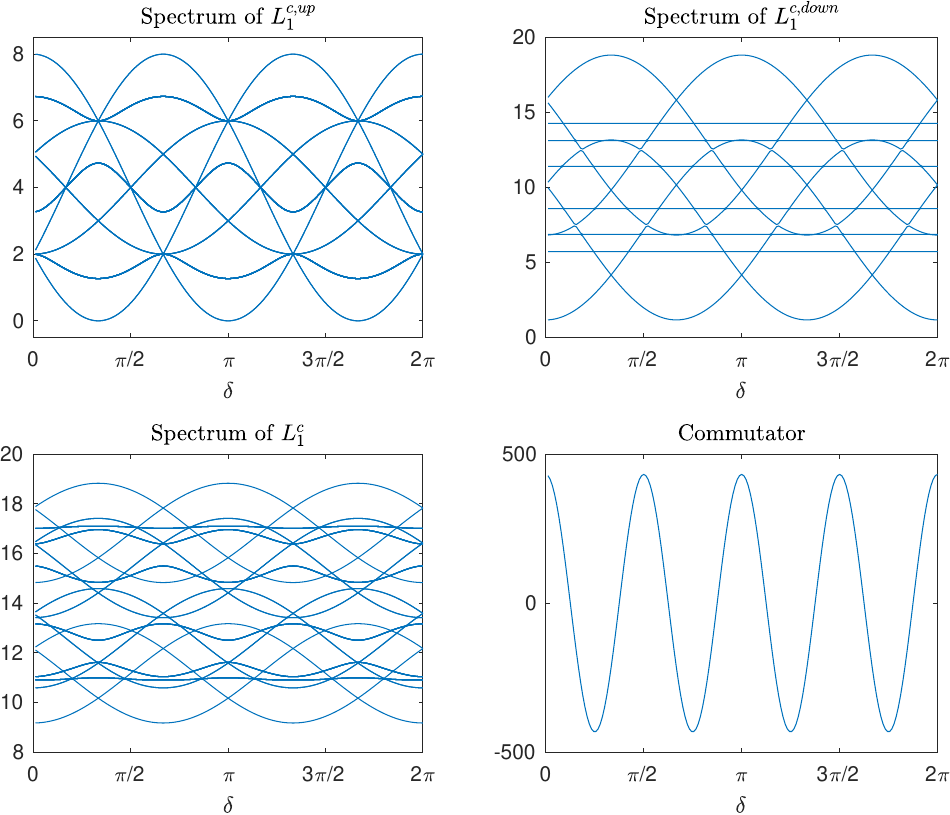}
\caption{The complete spectra of the $1$-Connection Laplacians $\hL_{1}^{c, up}$ (top-left), $\hL_{1}^{c, down}$ (top-right), and $\hL_{1}^{c}$ (bottom-left), and commutator $[\hL_{1}^{c, up}, \hL_{1}^{c, down}] $(bottom-right) for the Type $1$ Triangulated Torus of size $3\times 3$ and $25$ edges shown in Figure $\ref{fig:torus1}$ are plotted as a function of $\delta$. Note that some eigenvalues are degenerate. }
\label{fig:torus1_spectrum}
\end{figure}

\begin{figure}
\centering
\includegraphics[width=0.9\textwidth]{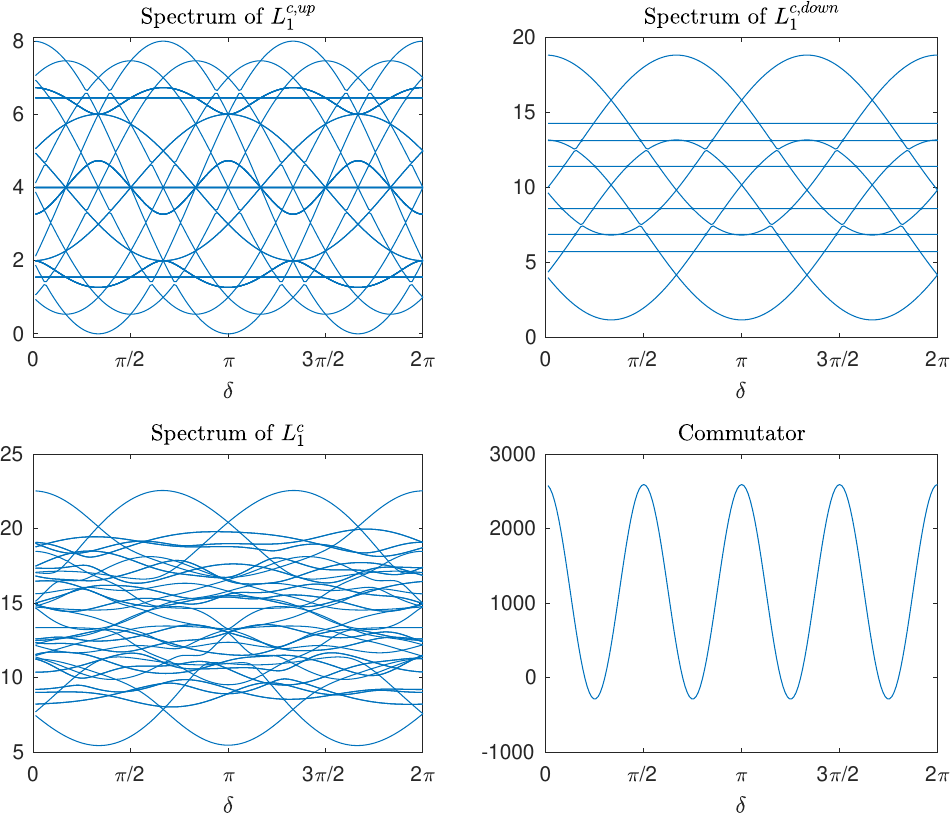}
\caption{The complete spectra of the $1$-Connection Laplacians $\hL_{1}^{c, up}$ (top-left), $\hL_{1}^{c, down}$ (top-right), and $\hL_{1}^{c}$ (bottom-left), and commutator $[\hL_{1}^{c, up}, \hL_{1}^{c, down}] $(bottom-right) for the Type $2$ Triangulated Torus of size $3\times 3$ and $25$ edges shown in Figure $\ref{fig:torus2}$ are plotted as a function of $\delta$. Note that some eigenvalues are degenerate but many degeneracies observed in the spectrum of the  Type 1 Triangulated Torus are lifted due to the presence of the frustration induced by the directions of the triangles.}
\label{fig:torus2_spec}
\end{figure}

The effect induced by the frustration of the flow direction is apparent from the comparison among the spectrum of the $1$-Connection Laplacians of  Type 1 and Type 2 Triangulated Tori (see Figure $\ref{fig:torus1_spectrum}$ and Figure $\ref{fig:torus2_spec}$).
Indeed for the Type $1$ Triangulated Torus it is clearly noticeable that the spectrum displays eigenvalues with much more significant degeneracy than for the Type $2$ Triangulated Torus case. Hence in the Type $2$ Triangulated Torus, the presence of the frustration induced by the triangle directions lifts the degeneracy of multiple eigenvalues, reflecting a decrease in the symmetries of these directed simplicial complexes. 





\section{Conclusion}
\label{chp3:conclusion}

Directed simplicial complexes constitute an important challenge in network theory as the directionality of interactions is ubiquitous in complex systems. Yet there is not a fully developed mathematical framework for treating directed simplicial complexes. As a matter of fact, so far there is not even consensus on the most suitable definition of directed simplicial complexes.
Here we tackle this challenge by leveraging the popular Magnetic Laplacian to treat directed graphs and networks.
We show that to formulate corresponding Hermitian operators that can capture all the possible configurations induced by the relative directions of simplices we may consider Higher-order Connection Laplacians making use of the Pauli matrices and of an additional rotation in the complex plane. Specifically, we built the $0$-up, $1$-up, $1$-down, and $2$-up Connection Laplacians of tesselations of $2$ dimensional orientable manifolds, where the $1$-up and $1$-down Connection Laplacians can be defined on arbitrary simplicial complexes of any dimensions.
The higher-order Connection Laplacians are used to formulate higher-order diffusion dynamics that can capture the frustration induced by incoherent directions of incident simplices. The application of the framework is investigated on simple and instructive examples of $2$-dimensional simplicial complexes.
In conclusion this work provides a framework for defining and treating directional simplicial complexes and higher-order diffusion dynamics enfolding on them. We build on the increasingly popular idea of adopting complex valued weights to combine dynamical processes on graphs and networks with non-trivial algebraic operations. We hope that this work will be useful for network scientists and applied mathematicians focusing on higher-order networks, complex weights and the interplay between dynamical processes and algebraic operations.

\ack{This project was carried out during X.G.'s Alan Turing Institute Enrichment Scheme Placement, supported by the Alan Turing Institute. X.G. acknowledges support of  MAC-MIGS CDT under EPSRC grant EP/S023291/1. D.J.H. was supported by EPSRC grant EP/P020720/1. K.C.Z. was supported by the Leverhulme Trust grant  2020-310 and EPSRC grant EP/V006177/1. }

\section*{Data, code, and materials}
Code for the experiments is available at \url{https://github.com/XueGong-git/Higher-order-Connection-Laplacians}.

\section*{Competing interests}
The authors declare that they have no conflicts of interest. 

\section*{References}
\bibliographystyle{vancouver}
\bibliography{references}
\end{document}